\documentclass[reprint,aps,pre]{revtex4-2}
\usepackage{graphicx}
\usepackage{amssymb}
\usepackage{amsmath}
\usepackage{gensymb}
\usepackage{xcolor}
\usepackage{hyperref}
\usepackage[utf8]{inputenc}
\usepackage{comment}
\usepackage{lipsum}
\usepackage[normalem]{ulem}
\usepackage{orcidlink}
\usepackage{subfigure}
\usepackage[section]{placeins}
\usepackage[margin=1.5cm]{geometry}
\begin{document}
\title{Finite-size effects and interaction-driven crossovers in quarter-filled attractive Hubbard model: Exact diagonalization, DMRG and machine-learning analysis}
\author{Md Fahad Equbal$^1$ \orcidlink{0009-0004-0054-9068}}
\email{mfequbal33@gmail.com (Corresponding Author)}
\author{M. A. H. Ahsan$^1$ \orcidlink{0000-0002-9870-2769}}
 \email{mahsan@jmi.ac.in}
\author{Satoru Hayami$^2$ \orcidlink{0000-0001-9186-6958}}
\email{hayami@sci.hokudai.ac.jp}
\affiliation{$^{1}$Department of Physics, Jamia Millia Islamia (Central University), New Delhi $110025$, India}
\affiliation{$^{2}$Department of Physics, Hokkaido University, Sapporo 060-0810, Japan}
\date{\today}
\begin{abstract}
We investigate the quarter-filled attractive Hubbard model on finite-width cylindrical lattices using exact diagonalization (ED), density-matrix renormalization group (DMRG) and unsupervised machine-learning-based techniques. Analysis of the ground-state energetics, local observables and correlation functions reveals a continuous interaction-driven crossover from weakly correlated fermions to a regime dominated by tightly bound singlet pairs. This crossover originates from the competition between kinetic-energy-driven fermionic itinerancy and interaction-driven onsite pair formation and exhibits behavior consistent with the BCS--BEC crossover in the thermodynamic limit. Hole-binding-energy calculations provide direct energetic evidence for pair formation: the two-hole binding energy remains negative throughout the attractive regime whereas three-hole binding emerges only at sufficiently strong attraction and exhibits pronounced finite-size dependence. To obtain an unbiased characterization of the correlation landscape, we apply principal component analysis (PCA) and uniform manifold approximation and projection (UMAP) to the real-space correlation matrices. PCA reveals a systematic redistribution of correlation variance whereas UMAP identifies a clear separation between weak- and strong-pairing regimes. Both machine-learning-based approaches independently identify the same crossover region inferred from conventional observables while providing an order-parameter-independent characterization of the underlying reorganization of many-body correlations. Finite-size scaling analyses of the pairing structure factor and the leading PCA variance ratio demonstrate that these signatures remain robust with increasing system size. Our results establish a consistent picture of interaction-driven pairing in the quarter-filled attractive Hubbard model and demonstrate the effectiveness of combining many-body numerical methods with unsupervised machine learning techniques for the analysis of correlated quantum systems. 

\end{abstract} 
\maketitle
\section{Introduction} 
\label{Intro}
The Hubbard model \cite{Hubbard1963} remains one of the central paradigms for understanding strongly correlated electron systems \cite{Dagotto1994,Daniel2022}. Originally introduced to describe electron localization and magnetism in narrow-band materials \cite{Mott1974,Gutzwiller1963,Kanamori1963}, it continues to serve as a minimal framework for investigating competing electronic orders, unconventional superconductivity and quantum phase transitions \cite{Kaito2022, Fahadboh2025, Fahadalm2026}. Among its various realizations, the attractive (negative-$U$) Hubbard model (AHM) \cite{Micnas1990} has attracted sustained interest because it captures the essential physics of fermionic pairing, superconductivity, charge ordering and the crossover between weakly bound Cooper pairs and tightly bound composite bosons \cite{Chen2024}. In recent years, renewed attention has emerged from studies of pair-density-wave states \cite{Chan2020,Zhu2025}, nonequilibrium pairing phenomena \cite{He2025} and the realization of controllable quantum simulators based on ultracold atoms in optical lattices \cite{Ho2009,Titas2025}.

Away from half filling, the AHM exhibits a rich interplay between pairing correlations, charge fluctuations and finite-density effects. In particular, quarter filling provides an especially interesting regime where particle-hole symmetry is absent and the competition between mobile Cooper-like pairs, charge-ordering tendencies and phase separation becomes more subtle \cite{Scalettar1989,Kaneko2014,Rodrigo2022,Tong2024}. A central question concerns how the nature of pairing evolves as the attractive interaction strength increases from the weak-coupling BCS regime to the strong-coupling limit of tightly bound local pairs. While this evolution is generally understood as a smooth BCS--BEC-like crossover rather than a true quantum phase transition \cite{Micnas1990,Kaneko2014,Seher2018,Rodrigo2022,Fahad2025}, the manifestation of this crossover in finite geometries and its dependence on system size remain incompletely understood.

Understanding finite-size effects is particularly important because both numerically exact and approximate many-body methods are inherently limited to finite clusters or finite-width cylinders. Correlation patterns observed on small systems may evolve significantly as the lattice width increases toward the two-dimensional (2D) limit. Distinguishing genuine many-body phenomena from finite-size artifacts is therefore crucial for establishing reliable signatures of pairing and interaction-driven crossovers. Systematic investigations of finite-width effects are directly relevant to quantum-simulation experiments where finite and quasi-one-dimensional (1D) geometries are routinely realized. 

Experimentally, ultracold fermionic atoms trapped in optical lattices have emerged as a highly tunable platform for realizing Hubbard-type Hamiltonians \cite{Leticia2018}. Through Feshbach-resonance techniques, interaction strengths can be tuned continuously from weak to strong coupling enabling controlled exploration of the attractive Hubbard regime \cite{Cheng2010}. Recent quantum-gas-microscope experiments have achieved site-resolved measurements of spin, charge and pairing correlations \cite{Boll2016,Mark2025,Gross2017}, providing unprecedented opportunities to directly probe many-body correlation functions. Furthermore, alkaline-earth and multi-component atomic systems have opened additional routes toward realizing attractive lattice fermion models with high experimental control \cite{Masaya2024,Alexander2025}. These developments motivate theoretical studies capable of providing quantitative benchmarks and identifying robust signatures of interaction-driven pairing phenomena in experimentally accessible geometries.

From the theoretical perspective, the AHM has been investigated using a wide range of analytical and numerical techniques, including quantum Monte Carlo (QMC) \cite{Hirsch1985,White1989}, exact diagonalization (ED) \cite{Callaway1990,Chen2023}, dynamical mean-field theory (DMFT) \cite{Vijay2008, Toschi2005}, variational approaches \cite{Kaneko2014}, tensor-network methods and density matrix renormalization group (DMRG) calculations \cite{Jiang2019,Jiang2020,Jiang2021,Jiang2024}. These studies have established the enhancement of pairing correlations with increasing attraction and have provided important insights into superconducting and charge-ordered states. Nevertheless, systematic investigations of finite-width effects at quarter filling remain relatively limited, particularly those combining conventional many-body observables with modern machine-learning-based approaches capable of extracting information from the full correlation structure.

In this work, we investigate the quarter-filled AHM on cylindrical geometries and systematically examine the interplay between interaction strength, correlation effects and finite-size behavior. We consider lattices of size $L_y\times L_x$ with periodic boundary conditions along the $x$ direction and open boundary conditions along the $y$ direction. Here $L_y$ and $L_x$ are number of sites along $y$ and $x$ directions, respectively. Throughout this work we fix $L_x=4$ and vary the cylinder width from $L_y=2$ to $6$. The total number of lattice sites is $M=L_xL_y$ and quarter filling corresponds to an average density $n=N/M=0.5$ where $N=M/2$ is the number of electrons.

To address this problem, we employ multi-method framework combining ED, DMRG and unsupervised machine-learning techniques. ED provides numerically exact ground-state energies, excitation spectra and correlation functions for the $2\times4$, $3\times4$ and $4\times4$ clusters, enabling complete access to the many-body wave function and physical quantities of interest. DMRG calculations extend the analysis to larger cylinders ($5\times4$ and $6\times4$), allowing controlled finite-size scaling and systematic investigation of width-dependent effects. Together, these approaches provide highly accurate energetic and correlation benchmarks across a broad interaction range.

Beyond conventional order-parameter analysis, we analyze charge-charge and pair-pair correlation matrices using principal component analysis (PCA) \cite{Leiwang2016,Hu2017,Costa2017,Kiwata2019} and uniform manifold approximation and projection (UMAP) \cite{Leland2018,Joana2022}. Correlation matrices contain substantially more information than a small set of local observables but their high dimensionality often obscures the dominant physical trends. Dimensionality-reduction techniques provide a systematic framework for identifying hidden structures, clustering behavior and interaction-driven crossovers directly from the complete correlation data. Recent studies have demonstrated the effectiveness of such unsupervised-learning approaches in detecting phase transitions and emergent orders in correlated-electron systems without prior assumptions regarding the relevant order parameters \cite{Khatami2019,Fahadps2026,Fahadap2026}. A central objective of the present work is therefore to assess whether unsupervised machine-learning techniques can identify the interaction-driven crossover in the AHM directly from correlation data and whether they provide a more objective characterization of the evolving many-body state than conventional observables alone.

One of the main aspects of this work is to determine how finite-size effects influence the evolution from weakly correlated fermions to strongly bound pairs in the quarter-filled AHM and to establish whether interaction-driven crossovers can be consistently identified using both conventional many-body observables and machine-learning analysis of correlation functions. By combining ED, DMRG, PCA and UMAP within a unified framework, we obtain a comprehensive characterization of ground-state energetics, pairing tendencies, correlation structures and finite-width effects. Our results provide quantitative benchmarks for future optical-lattice experiments and demonstrate how modern unsupervised-learning methods can complement traditional many-body techniques in the study of strongly correlated quantum systems.
 
This paper is organized as follows. In Section \ref{model}, we introduce the AHM and outline the numerical and machine-learning-based methodologies employed in this work. Section \ref{resdis} presents the results and discussion, including ground-state energetics, local observables, correlation functions, hole-binding energies, PCA, UMAP embeddings and finite-size scaling. Finally, Section \ref{summary} summarizes the main conclusions and discusses their implications for interaction-driven crossovers and machine-learning-based characterization of correlated quantum systems.
\section{Model and Method}
\label{model}
The one-band Hubbard Hamiltonian \cite{Hubbard1963, Gutzwiller1963, Kanamori1963} in real space is written as 
\begin{equation}
H=-t\sum_{\langle ij\rangle\sigma}(c_{i\sigma}^\dagger c_{j\sigma}+c_{j\sigma}^\dagger c_{i\sigma})+U\sum_i n_{i\uparrow}n_{i\downarrow},
\label{hamil}
\end{equation}
\noindent
where $c_{i\sigma}^\dagger (c_{i\sigma})$ is the fermionic operator that creates (annihilates) an electron with spin $\sigma \in \{\uparrow, \downarrow\}$ at lattice site $i$, and $\langle ij\rangle$ denotes nearest-neighbor (NN) sites on the lattice. The parameters $t$ and $U$ represent the NN hopping matrix amplitude and the on-site Coulomb interaction, respectively. The number operator $n_{i\sigma} = c_{i\sigma}^\dagger c_{i\sigma}$ counts particles at site $i$ with spin $\sigma$. We set $t=1$ as the unit of energy.     

We employ ED to compute ground and low-lying excited states on finite-size clusters. ED provides numerically exact access to eigenvalues and arbitrary correlation functions, making it an ideal platform for exploring both conventional observables and emerging machine-learning-based diagnostics. To extend the accessible system size, we exploit full spin-rotational symmetry and construct spin-adapted basis states \cite{Sarmahsan1996}. This symmetry adaptation block-diagonalizes the Hamiltonian into fixed-spin sectors and significantly reduces the effective Hilbert-space dimensionality, allowing us to investigate specific spin manifolds with high numerical precision.

While ED is restricted to small lattices by the exponential growth of the Hilbert space dimensionality, small Hubbard clusters remain valuable theoretical laboratories: they retain the full local interaction structure of the model, capture the short-range charge, spin and pairing correlations and permit controlled exploration of geometry and boundary conditions dependence. This is particularly relevant in light of recent discussions on the representability of finite-cluster results for Hubbard model physics and the increasing experimental realization of engineered small correlated systems in quantum simulators and ultracold atoms.

To access larger system sizes beyond the reach of ED, we employ the DMRG method \cite{Alvarez2009,Alvarez2011,Alvarez2012,Alvarez2013}, which provides highly accurate approximations to the ground state of strongly correlated lattice models in low-dimensional geometries. The DMRG calculations are performed on cylindrical lattices with periodic boundary conditions along the $x$ direction and open boundary conditions along the $y$ direction, fixing $L_x=4$ and considering widths up to $L_y=6$. 
We use the two-site finite-system algorithm with a maximum bond dimension of up to 1200 depending on system size and interaction strength. A truncation tolerance of $10^{-8}$ is employed together with ten finite-system sweeps following the infinite-system warmup procedure. The ground-state energies are monitored throughout the sweeps and found to converge to the reported precision. These convergence parameters ensure the numerical reliability of the observables and finite-size extrapolations presented below. 
The simulations target the ground state in a fixed particle-number sector corresponding to quarter filling. For each interaction strength $U$, we compute the ground-state energy, local observables and two-point correlation functions.

To characterize pairing tendencies beyond local observables, we further compute two-, three- and four-hole binding energies. These quantities provide a direct measure of the stability of multi-particle bound states and allow us to distinguish between simple pair formation and larger cluster formation at strong attraction. The DMRG results therefore complement the ED calculations by enabling systematic finite-size scaling and providing access to larger cylinders where finite-size effects are reduced.

To investigate whether the evolution of correlations with interaction strength contains hidden low-dimensional structure, we employ PCA, one of the most widely used linear dimensionality-reduction techniques in condensed matter applications \cite{Leiwang2016,Hu2017,Costa2017,Kiwata2019}.

For each value of $U$, the independent elements of the correlation matrices are arranged into a feature vector
\begin{equation}
\mathbf{x}(U)=(G_{12},G_{13},\cdots,G_{MN}),
\end{equation}
where $G_{ij}$ denotes the elements of the correlation matrix. The resulting dataset is standardized by subtracting the mean and dividing by the standard deviation of each feature before constructing the covariance matrix
\begin{equation}
\Sigma=\frac{1}{N_s-1}Q^{T}Q,
\end{equation}
where $N_s$ is the number of samples and $Q$ is the standardized data matrix. Diagonalization of $\Sigma$ yields principal components ordered according to their explained variance. Projection onto the leading principal components provides a low-dimensional representation of the correlation landscape and enables visualization of interaction-driven evolution across the attractive regime. In the present work, PCA serves as a linear probe for identifying dominant correlation channels and possible crossover regions associated with changes in the structure of many-body correlations.

While PCA captures only linear relationships among observables, correlation functions of interacting quantum systems often lie on nonlinear manifolds. To uncover such structures, we employ UMAP, a nonlinear dimensionality-reduction technique based on manifold learning and topological data analysis \cite{Leland2018}.
UMAP constructs a weighted graph in the high-dimensional feature space using NN relationships and seeks a low-dimensional embedding that preserves the local connectivity structure of the original dataset. Given a set of correlation vectors $\mathbf{x}(U)$, the algorithm first constructs a fuzzy simplicial complex describing local neighborhoods and then optimizes a low-dimensional representation by minimizing the discrepancy between the high- and low-dimensional graphs \cite{Leland2018}.

In this work, UMAP is applied to the correlation-matrix datasets derived from the many-body ground states. The resulting embeddings provide an unsupervised visualization of the interaction-driven evolution of correlation patterns across the attractive regime. Unlike PCA, UMAP can reveal nonlinear organization of the data and often separates different correlation regimes more clearly. Consequently, it serves as a complementary tool for identifying crossover regions, assessing finite-size effects and quantifying similarities among correlation patterns across different interaction strengths.
\section{Results and Discussion}
\label{resdis}
We now present our results on the quarter-filled AHM. Our analysis combines conventional many-body observables with unsupervised machine-learning techniques to investigate the interplay between interaction strength, pairing correlations and finite-size effects. We first discuss ground-state energetics, local observables and two-point correlation functions. We then examine hole-binding energies to assess pairing stability. Subsequently, we apply PCA and UMAP to the correlation matrices to identify dominant correlation patterns and interaction-driven crossovers. Finally, we perform finite-size scaling to evaluate the persistence of these features with increasing cylinder width.

\subsection{Ground-State Properties and Correlation Effects}
\label{gspce}
\subsubsection{Ground-state energy}
\begin{figure}[h]
\centering
\includegraphics[scale=0.60]{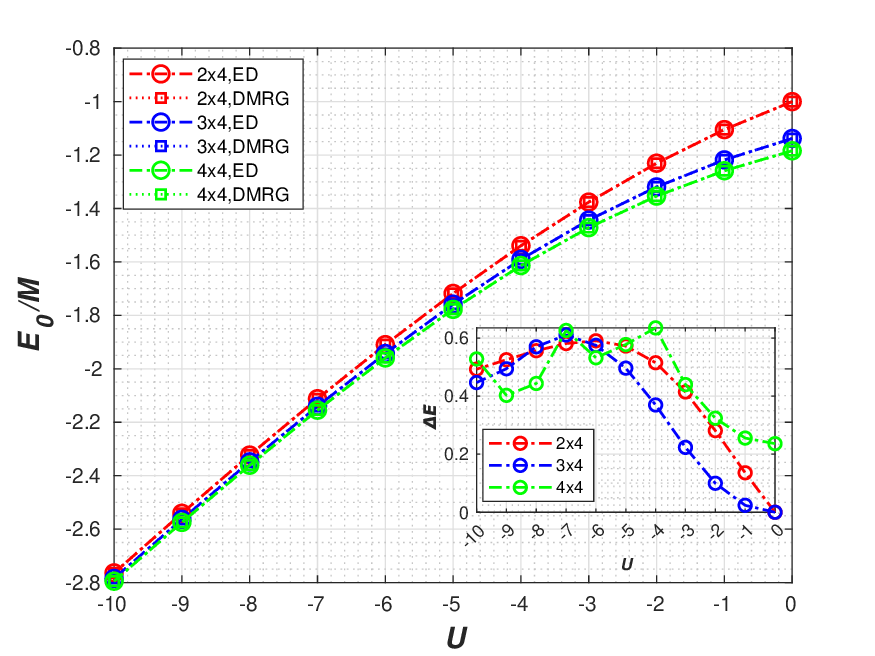}
\caption{Ground-state energy per site $E_0/M$ as a function of $U$ for $L_y\times4$ cylinders ($L_y=2,3,4$) at quarter filling, computed using ED (circle) and DMRG (square). Excellent agreement between ED and DMRG is observed. Inset: Excitation gap $\Delta E = E_1 - E_0$ obtained from ED. The broad maximum of $\Delta E$ at intermediate attraction signals a crossover from weakly bound fermions to strongly paired regime.}
\label{fig:Fig1}
\end{figure}

Figure \ref{fig:Fig1} shows the ground-state energy per site $E_0/M$ as a function of $U$ for quarter-filled $L_y\times4$ cylinders with $L_y=2,3$ and $4$. The ED and DMRG results are in excellent agreement over the entire interaction range, providing a stringent validation of the DMRG calculations and establishing a reliable basis for the subsequent analysis of larger systems.
As the attractive interaction increases, the ground-state energy decreases monotonically, reflecting the progressive enhancement of attractive pairing correlations. Finite-width effects are visible but relatively weak, with narrower cylinder ($L_y=2$) yielding slightly higher energies.

The inset displays the excitation gap $\Delta E=E_1-E_0$ obtained from ED. For all system sizes, $\Delta E$ exhibits a non-monotonic dependence on $U$, increasing from weak attraction, reaching a broad maximum around $U\approx -6$ to $-7$, and subsequently decreasing in the strongly attractive regime. This behavior signals an interaction-driven crossover separating a weak-coupling regime of Cooper-like pairs from a strong-coupling regime dominated by tightly bound local pairs. The location of the gap maximum therefore provides an estimate of the crossover region, consistent with the expected finite-size signatures of the smooth BCS--BEC-like evolution in the AHM away from half filling \cite{Micnas1990,Kaneko2014,Rodrigo2022,Fahad2025}.

Physically, this crossover originates from the competition between the electronic kinetic energy and the attractive interaction. For weak attraction, the kinetic energy dominates and the electrons remain largely itinerant forming spatially extended Cooper-like pairs. As $U$ becomes comparable to the bandwidth, the gain in interaction energy from onsite singlet formation competes strongly with the kinetic energy leading to a rapid reorganization of the low-energy states. Beyond this regime, tightly bound local pairs are formed and the relevant low-energy degrees of freedom evolve from fermionic quasiparticles to composite bosonic pairs whose effective mobility scales as $t_{\rm pair}\sim t^2/U$. The broad maximum of the excitation gap therefore marks the finite-size signature of the crossover between BCS-like and BEC-like pairing regimes expected in the thermodynamic limit.
\subsubsection{Local observables: double occupancy and local moment}
To gain further insight into the local electronic structure of the ground state, we next examine the average double occupancy and local moment, which provide direct measures of pair formation and local spin fluctuations, respectively. In the AHM, increasing negative $U$ favors the formation of onsite singlet pairs, leading to enhanced double occupancy and a corresponding suppression of local magnetic moments. We define the average double occupancy and local moment as \cite{Callaway1990}
\begin{eqnarray}
\bar{d}=\frac{M}{N^2}\sum_i\langle n_{i\uparrow}n_{i\downarrow}\rangle,
\\
\bar{m}=\frac{4}{N}\sum_i\langle (n_{i\uparrow}-n_{i\downarrow})^2\rangle,
\end{eqnarray}
where $M$ and $N$ denote the total number of lattice sites and electrons, respectively. 
\begin{figure}[h]
\centering
\includegraphics[scale=0.60]{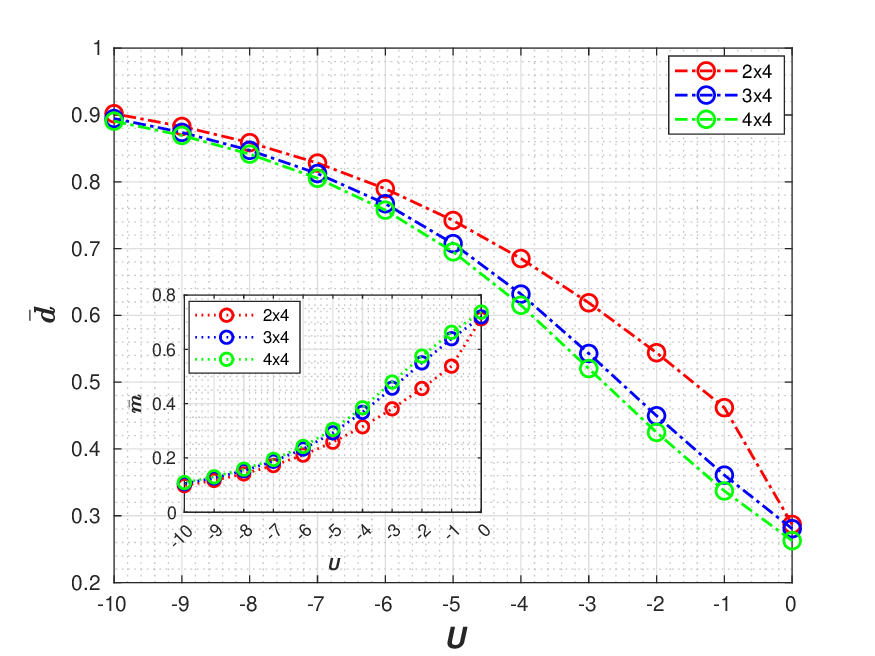}
\caption{Average double occupancy $\bar d$ as a function of $U$ for quarter-filled $L_y\times4$ cylinders ($L_y=2,3,4$). Inset: average local moment $\bar m$. Increasing attraction enhances onsite pair formation, resulting in a monotonic increase of $\bar d$ and a corresponding suppression of $\bar m$.}
\label{fig:Fig2}
\end{figure}

In Fig. \ref{fig:Fig2}, we present the average double occupancy $\bar d$ as a function of $U$. For all system sizes, $\bar d$ increases monotonically as the attractive interaction becomes stronger, indicating the progressive formation of onsite singlet pairs. In the weak-coupling regime, double occupancy remains small because electrons retain substantial kinetic mobility. As negative $U$ increases, opposite-spin electrons increasingly occupy the same site to gain interaction energy, leading to a rapid enhancement of $\bar d$. The growth becomes particularly pronounced in the intermediate-coupling region, consistent with the crossover regime identified from the excitation-gap analysis in Fig. \ref{fig:Fig1}. At strong attraction, $\bar d$ approaches a large value reflecting the dominance of tightly bound local pairs.

The inset shows the corresponding average local moment $\bar m$. In contrast to $\bar d$, the local moment decreases monotonically with increasing attraction reflecting the suppression of unpaired spins as electrons bind into onsite singlets. The complementary evolution of $\bar d$ and $\bar m$ demonstrates that the attractive interaction continuously converts local magnetic degrees of freedom into paired charge degrees of freedom. The weak dependence on cylinder width further indicates that pair formation is a robust local phenomenon, whereas finite-size effects primarily influence long-range correlations discussed below.
\subsubsection{Charge spin and pairing correlations}
While local observables provide information about onsite pair formation, the nature of the emerging many-body state is more revealed through nonlocal correlation functions. To characterize the competing charge, spin and superconducting fluctuations, we evaluate the charge-charge, spin-spin and pair-pair correlation functions defined as
\begin{eqnarray}
D_{ij}&=&\langle G\mid n_i n_j\mid G\rangle,
\\
L_{ij}&=&\frac{1}{4}\langle G\mid (n_{i\uparrow}-n_{i\downarrow})
(n_{j\uparrow}-n_{j\downarrow}) \mid G\rangle,
\\
P_{ij}&=&
\langle G\mid
c_{i\uparrow}^\dagger c_{i\downarrow}^\dagger
c_{j\downarrow} c_{j\uparrow}\mid G\rangle,
\end{eqnarray}
where $\mid G\rangle$ denotes the many-body ground state and $n_i=n_{i\uparrow}+n_{i\downarrow}$.
As the system size increases, the real-space correlation matrices become increasingly complex. It is therefore convenient to characterize their dominant ordering tendencies through the corresponding structure factors,
\begin{equation}
S_X(\mathbf q)=\frac{1}{M}\sum_{ij}e^{i\mathbf q\cdot(\mathbf R_i-\mathbf R_j)}X_{ij},
\end{equation}
where $X\in\{D,L,P\}$. Throughout this work, we focus on $\mathbf q=(\pi,\pi)$ for charge and spin correlations and $\mathbf q=(0,0)$ for uniform singlet pairing correlations.
\begin{figure}[h]
\centering
\includegraphics[scale=0.60]{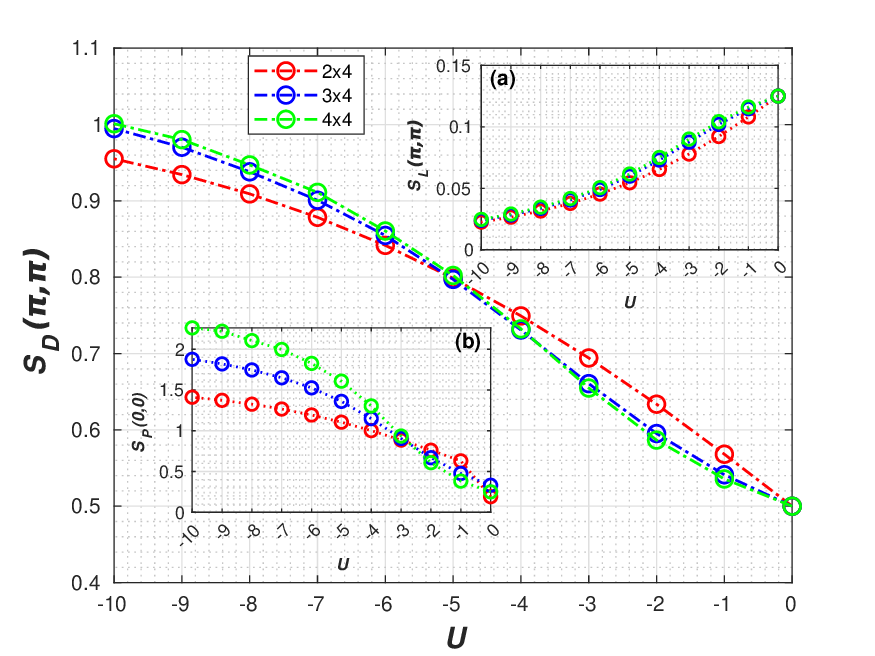}
\caption{Charge structure factor $S_D(\pi,\pi)$ as a function of $U$ for quarter-filled $L_y\times4$ cylinders ($L_y=2,3,4$). Inset (a): spin structure factor $S_L(\pi,\pi)$. Inset (b): singlet pairing structure factor $S_P(0,0)$. Increasing attractive interaction suppresses spin correlations while enhancing charge and pairing fluctuations.}
\label{fig:Fig3}
\end{figure}
Figure \ref{fig:Fig3} summarizes the evolution of the dominant correlation channels with increasing attractive interaction. The main panel shows the charge structure factor $S_D(\pi,\pi)$ which increases steadily as $U$ becomes more negative for all cylinder widths indicating a progressive enhancement of charge correlations associated with the formation of onsite pairs. The growth becomes more pronounced in the intermediate- and strong-coupling regimes, consistent with the crossover region identified from the excitation-gap and local-observable analyses.
The spin structure factor $S_L(\pi,\pi)$ shown in inset (a), exhibits the opposite trend. As the attraction increases, $S_L(\pi,\pi)$ is strongly suppressed, reflecting the depletion of local magnetic moments and the reduction of spin fluctuations due to singlet-pair formation. This behavior is fully consistent with the monotonic decrease of the average local moment observed in Fig. \ref{fig:Fig2}.
Inset (b) presents the singlet pairing structure factor $S_P(0,0)$. In contrast to the spin channel, $S_P(0,0)$ increases monotonically with increasing attraction, demonstrating the strengthening of superconducting correlations. The simultaneous enhancement of $S_D(\pi,\pi)$ and $S_P(0,0)$ together with the suppression of $S_L(\pi,\pi)$ reveals a gradual transfer of spectral weight from spin fluctuations to charge and pairing degrees of freedom. The relatively weak dependence on cylinder width further suggests that these trends are intrinsic features of the quarter-filled AHM rather than finite-size artifacts.

\begin{figure}[h]
\centering
\includegraphics[scale=0.40]{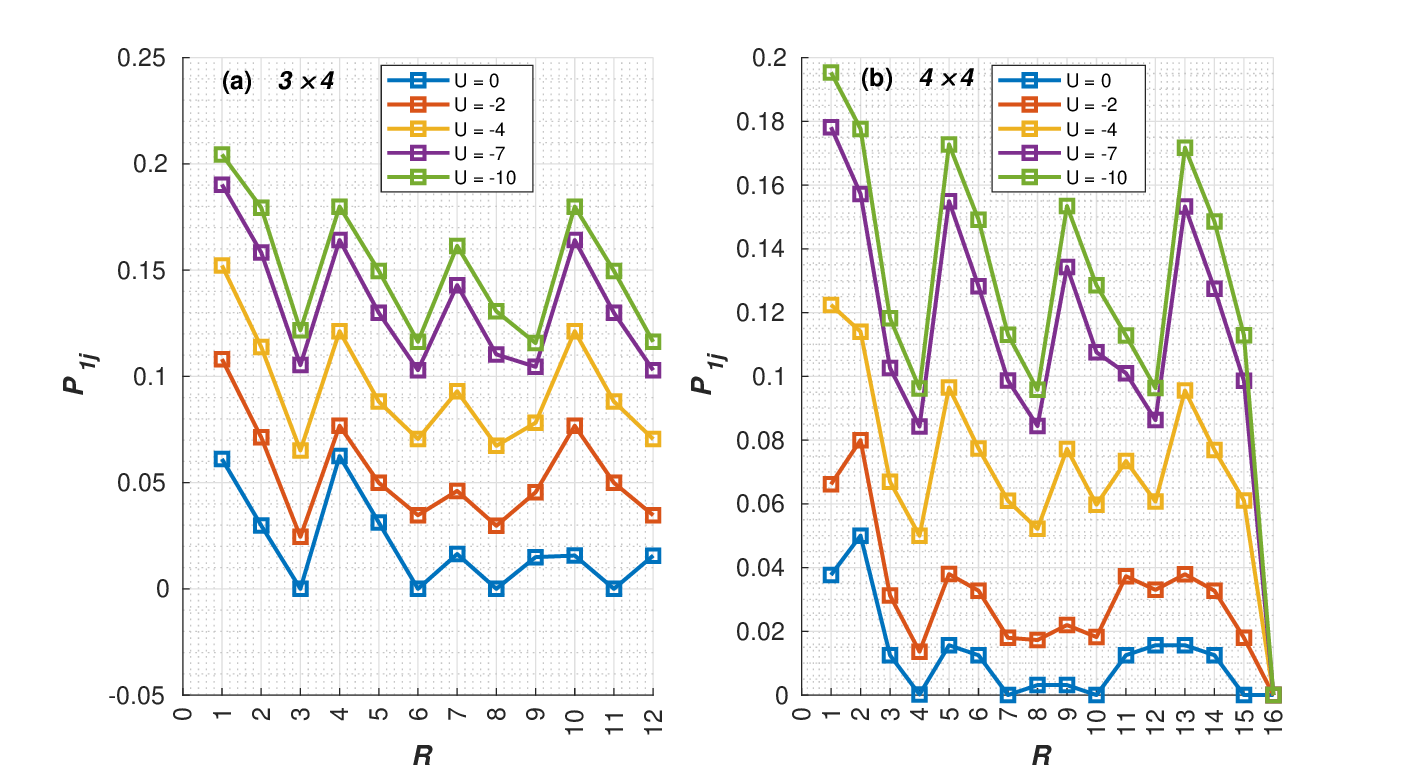}
\caption{Pair-pair correlation function $P_{1j}$ as a function of distance $R$ for (a) $3\times4$ and (b) $4\times4$ cylinders at quarter filling. Here site $1$ is chosen as the reference site. Increasing attractive interaction enhances both the magnitude and spatial extent of pairing correlations, indicating the progressive formation of bound singlet pairs.}
\label{fig:Fig4}
\end{figure}
To probe the spatial extent of superconducting correlations, we examine the real-space pair-pair correlation function $P_{1j}$ where site $1$ is chosen as a reference site and $j$ spans the remaining lattice sites. Figure \ref{fig:Fig4} shows $P_{1j}$ as a function of the distance $R$ for $3\times4$ and $4\times4$ cylinders at representative interaction strengths. For both system sizes, the magnitude of $P_{1j}$ increases systematically as the attractive interaction becomes stronger, demonstrating the progressive enhancement of singlet-pair correlations. 
For non-interacting ($U=0$) case, the pair-pair correlations are small and decay rapidly with distance, indicating limited pairing coherence.  At weak attractive interactions ($U=-2$ and $-4$), the amplitude of $P_{1j}$ grows substantially and remains finite over large separations. This enhancement reflects the increasing tendency of opposite-spin electrons to form bound singlet pairs and is consistent with the increase in double occupancy and pairing structure factors discussed previously.
For intermediate and strong attraction ($U=-7$ and $-10$), the correlations remain sizable throughout the cluster and exhibit oscillatory behavior arising from the finite lattice geometry and boundary conditions. The enhancement is particularly pronounced in the $3\times4$ cylinder, suggesting that finite-width effects can strengthen pairing correlations in narrower systems. Similar trends are observed for the $4\times4$ cluster, indicating that the growth of pairing correlations is robust across different system sizes.

These real-space results are consistent with the monotonic increase of the pairing structure factor shown in Fig. \ref{fig:Fig3} and provide microscopic evidence for the continuous evolution from weakly correlated fermions to a regime dominated by tightly bound local pairs, in agreement with the crossover identified from the excitation-gap and local-observable analyses.

\subsubsection{Binding energy of holes}
To establish whether the enhanced pairing correlations correspond to true bound states, we analyze the two-, three- and four-hole binding energies. Here, holes are added relative to the quarter-filled ground-state reference. We define the two-hole ($E_{B2}$), three-hole ($E_{B3}$) and four-hole ($E_{B4}$) binding energies as
\begin{equation}
E_{B2} = (E(2)-E(0))-2(E(1)-E(0)),
\end{equation}
\begin{equation}
E_{B3} = (E(3)-E(0))-(E(2)-E(0))-(E(1)-E(0)),
\end{equation}
\begin{equation}
E_{B4} = (E(4)-E(0))-2(E(2)-E(0)),
\end{equation}
where $E(h)$ denotes the ground-state energy of the system containing $h$ holes relative to the quarter-filled reference state $E(0)$. A negative value of $E_{B2}$ signifies a stable bound state of two holes (pairing), whereas negative values of $E_{B3}$ or $E_{B4}$ would indicate a tendency toward multi-hole clustering or phase separation.

\begin{figure}[h]
\centering
\includegraphics[scale=0.60]{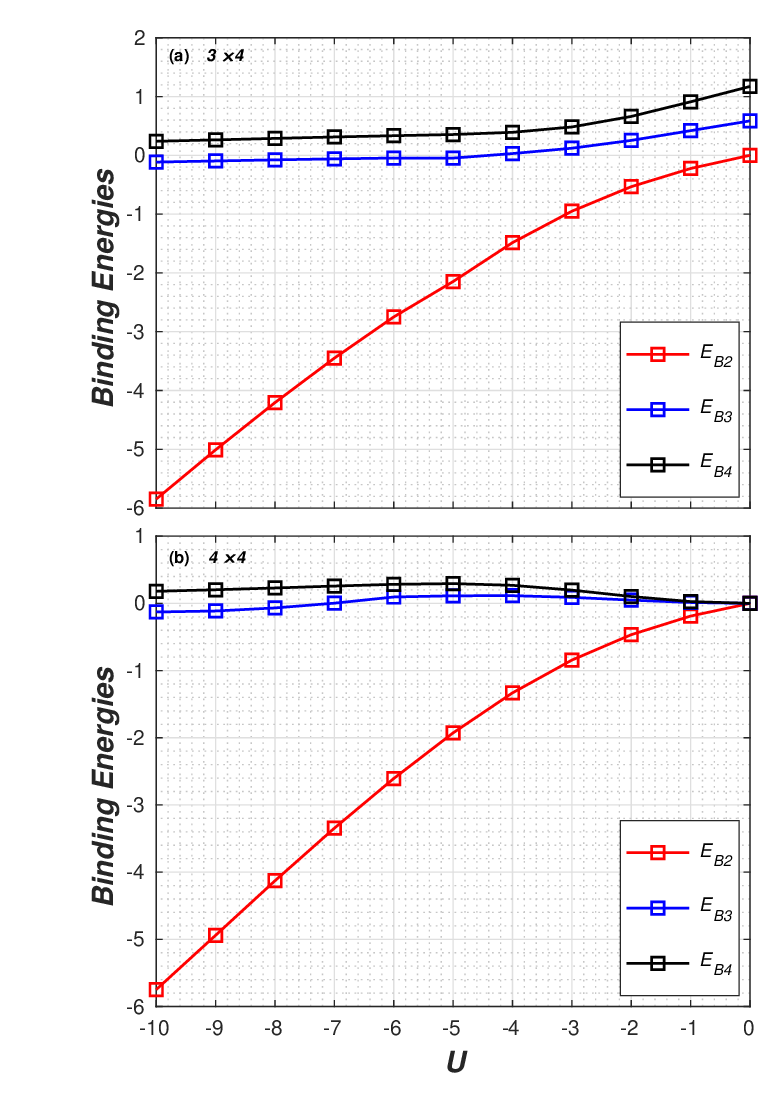}
\caption{Two-, three- and four-hole binding energies as functions of $U$ for (a) $3\times4$ and (b) $4\times4$ cylinders at quarter filling. The negative values of $E_{B2}$ demonstrate robust two-hole pairing throughout the attractive regime. The sign change of $E_{B3}$ at strong attractive interaction indicates finite-size-dependent three-hole clustering, while the positive values of $E_{B4}$ show that stable four-hole bound states do not form within the parameter range studied.}
\label{fig:Fig5}
\end{figure}
Figure \ref{fig:Fig5} shows the evolution of two-, three- and four-hole binding energies for $3\times4$ and $4\times4$ cylinders as functions $U$. For both system sizes, the two-hole binding energy $E_{B2}$ remains negative throughout the attractive regime and becomes progressively more negative with increasing negative $U$. This behavior demonstrates the robust formation of stable hole pairs and provides energetic confirmation of the enhanced pairing correlations observed in Fig. \ref{fig:Fig4}. As $U\rightarrow0$, $E_{B2}$ approaches zero, reflecting the disappearance of pair binding in the non-interacting limit.

The three-hole binding energy exhibits a pronounced finite-size dependence. For $3\times4$ cluster, $E_{B3}$ remains positive for $U>-5$ and becomes negative for $U\leq-5$ indicating that sufficiently strong attraction can stabilize three-hole clusters on the finite lattice. A similar trend is found for $4\times4$ cluster, although the sign change is shifted to stronger coupling, with $E_{B3}<0$ only for $U\leq-8$. The delayed onset of three-hole binding in the larger cluster suggests that the stability of such higher-order bound states is strongly influenced by finite-size effects and is considerably weaker than the tendency toward two-hole pairing.
In contrast, the four-hole binding energy $E_{B4}$ remains positive throughout the entire interaction range for both system sizes. This demonstrates that, although strong attraction can stabilize pairs and eventually induce three-hole clustering on finite clusters, there is no energetic evidence for the formation of stable four-hole bound states. The absence of negative $E_{B4}$ also argues against a tendency toward large-scale charge aggregation or phase separation within the parameter range studied.

Taken together, the binding-energy analysis reinforces the picture emerging from the local observables and correlation functions. The AHM at quarter filling is characterized by robust two-hole pairing across the entire attractive regime while larger bound clusters appear only at strong coupling and remain highly sensitive to finite-size effects. The persistent positivity of $E_{B4}$ further indicates that pairing, rather than extensive multi-particle clustering, constitutes the dominant low-energy instability of the system.
\subsection{Machine-Learning-Based Approaches}
\label{mlb}
While the conventional observables discussed above provide clear evidence for enhanced pairing correlation and an interaction-driven crossover, they probe only specific aspects of the many-body state. In contrast, the full correlation matrices contain considerably rich information regarding the collective organization of charge and pairing fluctuations. Extracting this information directly from high-dimensional correlation data can be challenging, particularly when multiple correlation channels evolve simultaneously with interaction strength.

To obtain an unbiased characterization of the interaction-driven evolution of the system, we employ unsupervised machine-learning techniques. In this section, we first apply PCA to identify the dominant fluctuation patterns encoded in the correlation matrices and quantify their relative importance. We then complement this linear analysis with the nonlinear dimensionality-reduction technique UMAP, which can reveal hidden manifold structures and interaction-driven clustering of correlation patterns that may not be captured by PCA alone.
\subsubsection{Principal component analysis}
\label{pca}
To obtain an unbiased, machine-learning-based characterization of how correlation patterns reorganize across the interaction-driven crossover, we apply PCA directly to the real-space correlation matrices introduced in the previous section. Unlike conventional structure factors, which project correlations onto predetermined momentum channels, PCA identifies the dominant collective fluctuation patterns solely from the statistical structure of the data, without assuming any specific order parameter or ordering wave vector.

For each interaction strength $U$, the independent elements of a correlation matrix are reshaped into a feature vector. The collection of vectors obtained for different interaction strengths forms a data matrix $Q$, whose columns correspond to correlation features and whose rows correspond to different values of $U$. Prior to analysis, each feature is centered by subtracting its mean value over the dataset.
The covariance matrix is then constructed as
\begin{equation}
\Sigma=\frac{1}{N_s-1}Q^{T}Q,
\end{equation}
where $N_s$ denotes the number of samples. Diagonalization of $\Sigma$ yields eigenvalues $\lambda_k$ and orthonormal eigenvectors $w_k$,
\begin{equation}
\Sigma w_k=\lambda_k w_k,
\end{equation}
where each eigenvector defines a principal component representing a collective fluctuation channel in the correlation space. The corresponding eigenvalue quantifies the amount of variance carried by that channel.

The projection scores onto the $k$-th principal component are given by
\begin{equation}
p_k=Qw_k,
\end{equation}
where each element of $p_k$ corresponds to a particular interaction strength.

To assess the relative importance of different fluctuation channels, we compute the normalized explained-variance ratio,
\begin{equation}
\tilde{\lambda}_k=\frac{\lambda_k}{\sum_i \lambda_i}.
\end{equation}
The leading variance ratios identify the dominant correlation patterns present in the data. By examining both the explained variance and the projections onto the leading principal components, one can track how the underlying correlation structure evolves with interaction strength in a fully unsupervised manner. PCA thus provides a compact low-dimensional representation of the many-body correlation landscape and offers a complementary perspective to the conventional observables discussed above. 

\begin{figure}[h]
\centering
\includegraphics[scale=0.41]{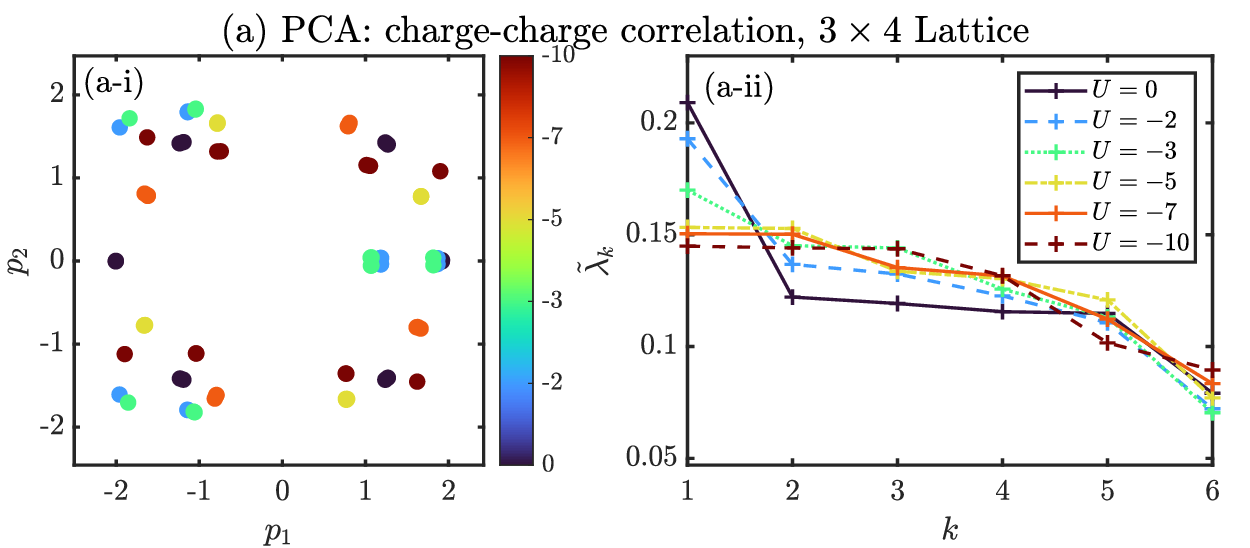}
\quad
\includegraphics[scale=0.41]{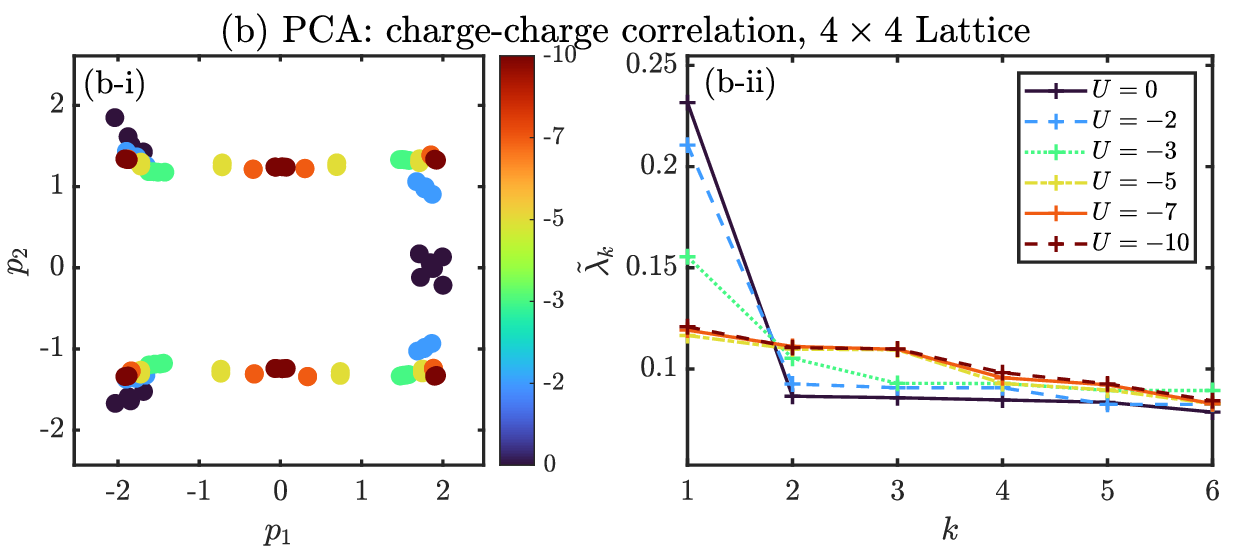}
\quad
\includegraphics[scale=0.41]{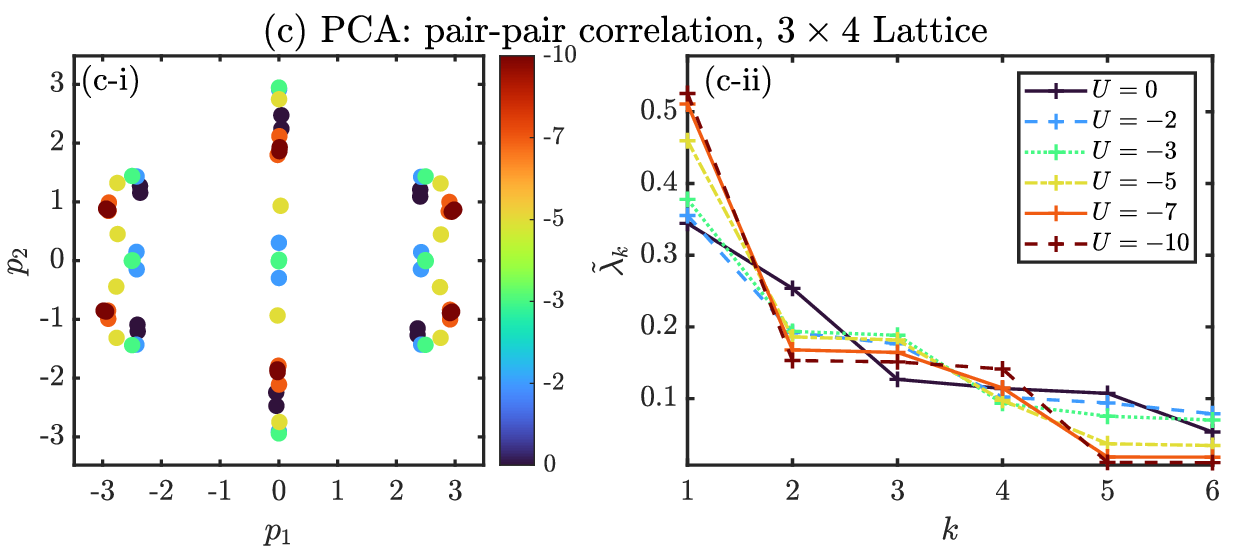}
\quad
\includegraphics[scale=0.41]{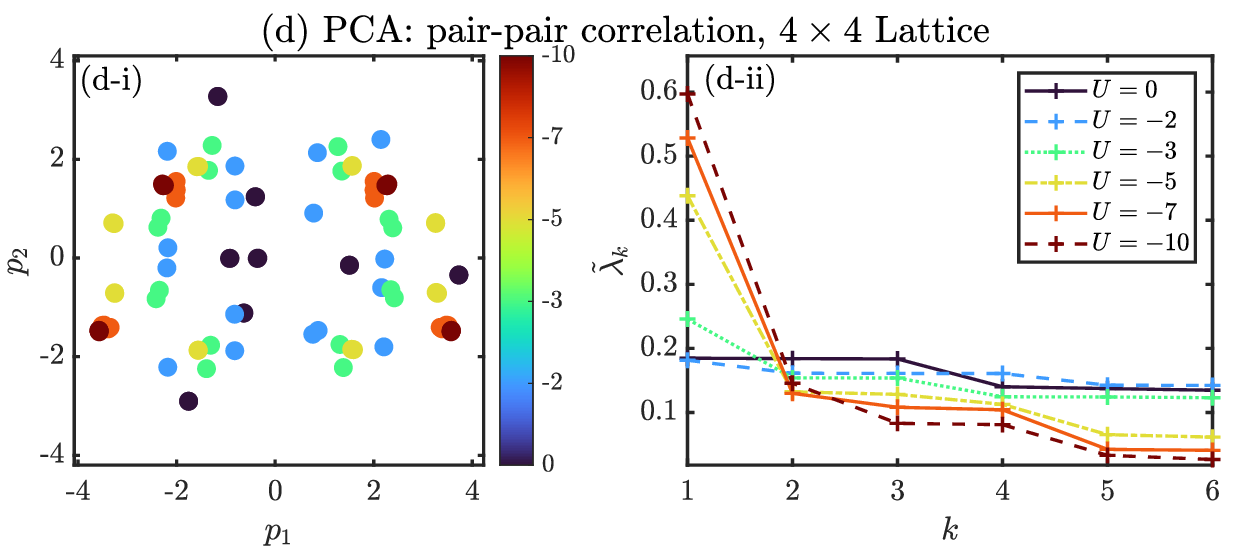}
\caption{Principal component analysis (PCA) of ground-state correlation matrices at quarter filling. Left columns show the projections onto the first two principal components ($p_1, p_2$), color-coded by the interaction strength $U$. Right columns show the normalized explained-variance ratio $\tilde{\lambda}_k$ for the first six components ($k=1$ to $6$). Panels correspond to: (a) charge-charge correlations on the $3 \times 4$ lattice, (b) charge-charge correlations on the $4 \times 4$ lattice, (c) pair-pair correlations on the $3 \times 4$ lattice and (d) pair-pair correlations on the $4 \times 4$ lattice.}
\label{fig:Fig6}
\end{figure}
Figure \ref{fig:Fig6} presents the PCA results obtained from the charge-charge and pair-pair correlation matrices for $3\times4$ and $4\times4$ lattices. The left panels show the projections onto the first two principal components $(p_1,p_2)$, while the right panels display the corresponding normalized explained-variance ratios $\tilde{\lambda}_k$. Together, these results provide an unsupervised characterization of how the many-body correlation landscape evolves with increasing attractive interaction.

The PCA projections evolve continuously with interaction strength for both lattice sizes, without exhibiting abrupt separation into distinct clusters. Instead, the data points are redistributed smoothly in the $(p_1,p_2)$ plane as $U$ becomes more negative, consistent with the interaction-driven crossover inferred from the excitation gap, local observables, pairing correlations and hole-binding energies.

More quantitative information is obtained from the explained-variance spectra. For the charge-charge correlation matrices [Figs. \ref{fig:Fig6}(a-ii) and \ref{fig:Fig6}(b-ii)], the leading explained-variance ratio $\tilde{\lambda}_1$ decreases systematically as the attractive interaction increases from $U=0$ to $U=-10$. In the weak-coupling regime ($U=0$ to approximately $-3$ or $-4$), the first principal component captures the largest fraction of the total variance, indicating that the charge-correlation data are dominated by a single fluctuation channel. As the attraction is strengthened ($U\lesssim-5$), however, the variance becomes more evenly distributed among the first few principal components, with two or more components carrying comparable weights. This redistribution indicates that no single fluctuation channel completely characterizes the charge-correlation data at strong attraction. Instead, several principal components contribute appreciably to the variance, reflecting the increasingly complex spatial organization of charge correlations associated with pair formation.

The pair-pair correlation matrices exhibit the opposite behavior. As shown in Figs. \ref{fig:Fig6}(c-ii) and \ref{fig:Fig6}(d-ii), the leading explained-variance ratio $\tilde{\lambda}_1$ increases with increasing attractive interaction. For weak attraction ($U=0$ to approximately $-3$ or $-4$), the variance is shared among several principal components, reflecting the relatively distributed nature of pairing fluctuations. Beyond the crossover region ($U\lesssim-5$), the first principal component becomes increasingly dominant, while the contributions from higher-order components are substantially reduced. This concentration of variance indicates that the pair-pair correlation matrices become progressively governed by a single dominant collective fluctuation pattern as tightly bound singlet pairs develop.

The contrasting evolution of the explained-variance spectra for the charge-charge and pair-pair correlations provides an important machine-learning signature of the interaction-driven crossover. While charge correlations evolve toward a more distributed, multi-component representation, pairing correlations become increasingly low-dimensional and dominated by a single principal fluctuation pattern. The same qualitative behavior is observed for both the $3\times4$ and $4\times4$ lattices, demonstrating that these trends are robust against finite-size effects.

An important advantage of the PCA analysis is that it identifies the crossover directly from the global structure of the correlation matrices without requiring the specification of a particular observable or order parameter. Conventional quantities such as the excitation gap, pairing structure factor and hole-binding energy each probe only a specific aspect of the many-body state. In contrast, PCA analyzes all correlation elements simultaneously and reveals how the dominant fluctuation channels reorganize across the interaction range. The redistribution of explained variance among the principal components therefore provides an objective measure of the evolving correlation landscape. The pronounced evolution of the leading variance ratios near $U\approx -4$ to $-6$ coincides with the crossover region identified from conventional observables demonstrating that the machine-learning-based approach captures the same underlying physics while avoiding prior assumptions regarding the relevant order parameter. The increasing dominance of the leading principal component in the pair-pair correlation matrices at strong attraction reflects the emergence of a coherent pairing-dominated fluctuation channel. This behavior is consistent with the crossover from weakly overlapping Cooper pairs to a regime of tightly bound local pairs and therefore provides an unsupervised signature of the underlying BCS--BEC evolution.
\subsubsection{Uniform manifold approximation and projection}
Although PCA successfully identifies the dominant linear fluctuation channels, it is inherently limited to linear projections of the high-dimensional correlation space. In interacting quantum systems, however, the evolution of many-body correlations may follow nonlinear manifolds that cannot be fully represented by linear combinations of the original variables. To uncover such nonlinear structures, we next employ UMAP, a manifold-learning algorithm that preserves the local neighborhood relationships among data points while constructing a low-dimensional embedding.

Using the same feature vectors employed in the PCA analysis, UMAP first constructs a weighted nearest-neighbor graph in the original correlation space and interprets it as a fuzzy topological representation of the underlying data manifold. A low-dimensional embedding is then obtained by minimizing the difference between the high-dimensional and low-dimensional neighborhood graphs through stochastic optimization \cite{Leland2018}. Unlike PCA, which maximizes the global variance captured by orthogonal directions, UMAP is designed to preserve the intrinsic local geometry of the dataset and can therefore reveal nonlinear organization and clustering of correlation patterns. Consequently, UMAP provides an independent and complementary probe of the interaction-driven evolution of the many-body correlation landscape.
\begin{figure}[h]
\centering
\includegraphics[scale=0.38]{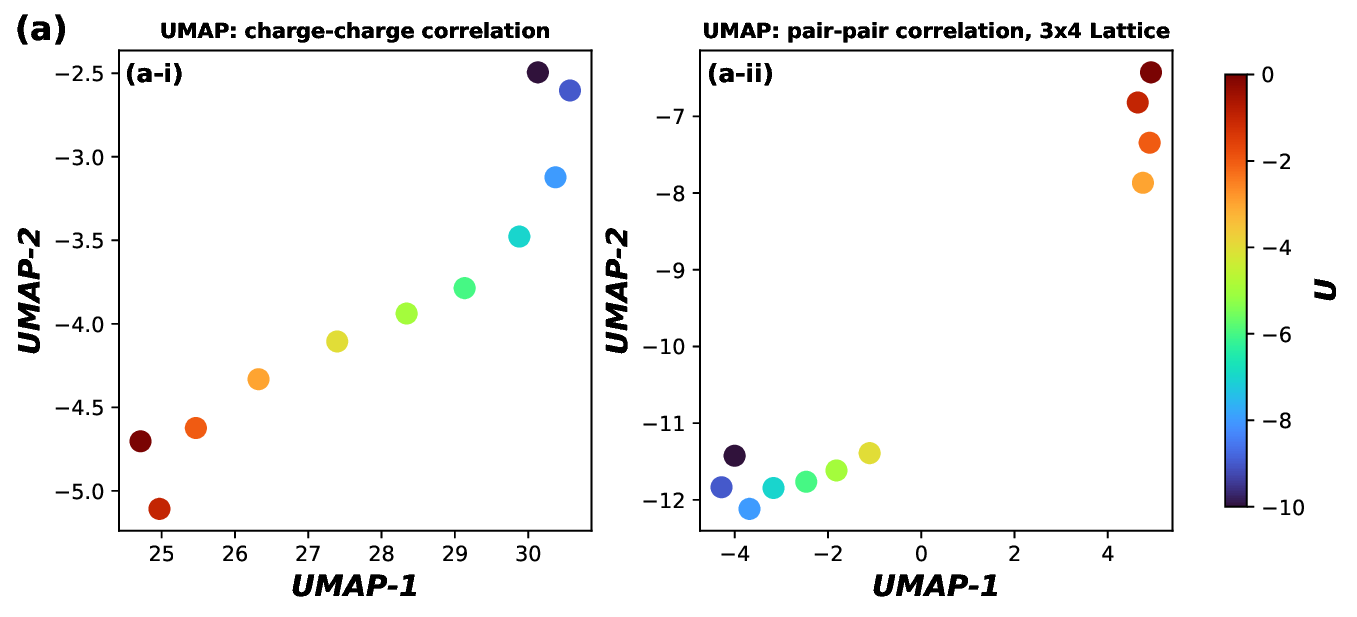}
\quad
\includegraphics[scale=0.38]{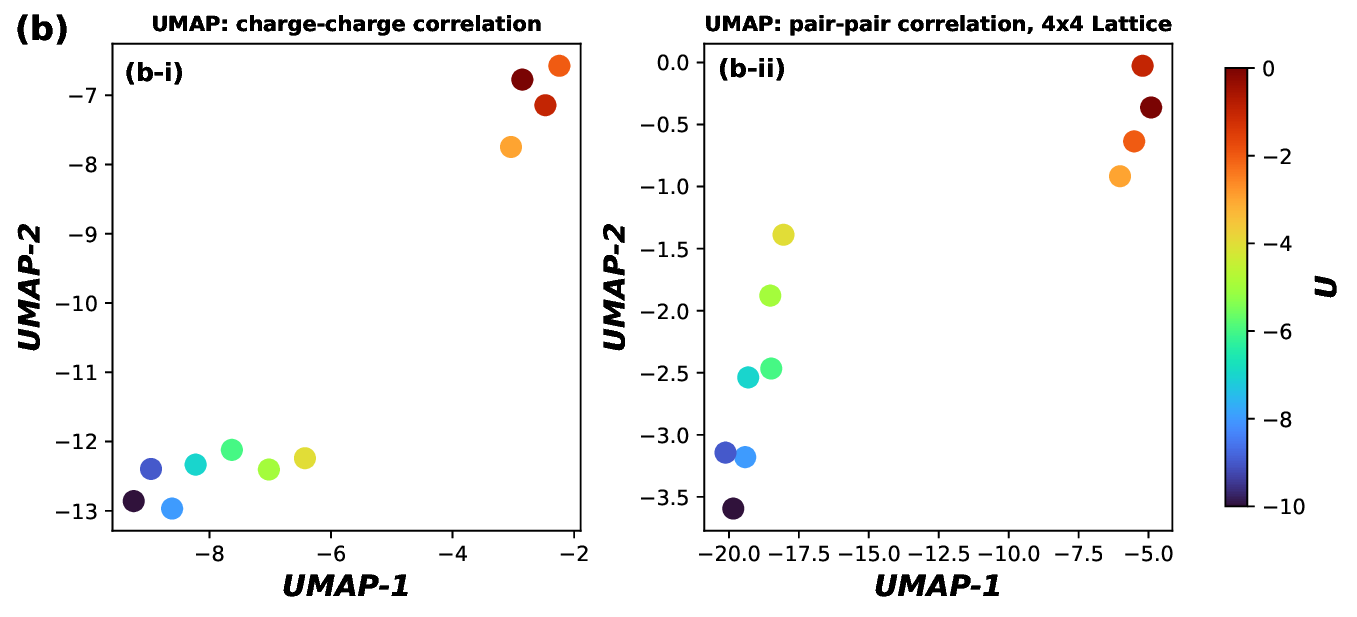}
\caption{UMAP embeddings of the ground-state correlation matrices at quarter filling. Colors denote the attractive interaction strength $U$. Panels (a-i) and (b-i) show the embeddings of the charge-charge correlation matrices for the $3\times4$ and $4\times4$ lattices, respectively, while (a-ii) and (b-ii) display the corresponding pair-pair correlation embeddings. The charge-correlation data evolve smoothly along a continuous manifold whereas the pair-correlation data separate into two distinct nonlinear branches providing a clear machine-learning-based signature of the interaction-driven crossover.}
\label{fig:Fig7}
\end{figure}

Figure \ref{fig:Fig7} presents the 2D UMAP embeddings obtained from the charge-charge and pair-pair correlation matrices for $3\times4$ and $4\times4$ lattices. Each point corresponds to a correlation matrix at a particular interaction strength with the color indicating the value of the attractive interaction $U$. For the charge-charge correlation matrices [Figs. \ref{fig:Fig7}(a-i) and \ref{fig:Fig7}(b-i)], the embeddings exhibit a systematic evolution with increasing attraction. In $3\times4$ lattice, the data points lie approximately along a continuous 1D manifold. For $4\times4$ lattice, the embedding develops two partially separated groups corresponding approximately to the weak- and strong-attraction regimes. However, the points remain ordered according to interaction strength and the separation emerges gradually rather than through an abrupt discontinuity. This behavior indicates a continuous reorganization of the charge-correlation patterns across the interaction range with the strongest changes occurring in the intermediate-coupling region. The same crossover scale is consistent with that identified from the conventional correlation functions and the PCA analysis suggesting that the UMAP embedding captures the evolving correlation landscape in a nonlinear and order-parameter-independent manner.

The pair-pair correlation matrices exhibit a markedly different behavior. As shown in Figs. \ref{fig:Fig7}(a-ii) and \ref{fig:Fig7}(b-ii), the UMAP embeddings separate into two clearly distinguishable branches corresponding approximately to the weak-coupling ($U=0$ to $-4$) and strong-coupling ($U\lesssim-5$) regimes. The transition between these branches occurs near the same interaction range identified previously from the excitation gap, real-space pairing correlations, hole-binding energies and PCA analysis. The emergence of two distinct nonlinear branches indicates that the structure of the pair-pair correlation matrices undergoes a substantial reorganization that is only partially captured by linear dimensionality reduction. The separation between the two branches becomes even more pronounced for the $4\times4$ lattice, indicating that the nonlinear distinction between weakly correlated fermions and the strongly paired regime becomes increasingly well resolved as the system size increases. Nevertheless, the qualitative organization of the embeddings remains the same for both lattices confirming that the observed nonlinear correlation structure is robust against finite-size effects.

Taken together, the UMAP analysis complements the PCA results by revealing the intrinsic nonlinear geometry of the correlation datasets. While the charge-correlation matrices evolve continuously along a single manifold, the pair-correlation matrices organize into distinct nonlinear branches associated with weak- and strong-pairing regimes. The crossover between these branches occurs in the same interaction range identified from the excitation gap, local observables, correlation functions and binding-energy analysis, providing an independent machine-learning-based confirmation of the interaction-driven crossover in the quarter-filled AHM. The separation into two nonlinear branches reflects the change in the intrinsic correlation structure of the many-body wavefunction as the system evolves from a regime of weakly bound overlapping Cooper pairs to one dominated by tightly bound local pairs.

Compared with conventional observables, the UMAP analysis provides additional insight into the geometry of the many-body correlation space. Whereas quantities such as the excitation gap or pairing structure factor indicate that a crossover occurs, UMAP visualizes how the full correlation matrices reorganize as the interaction strength changes. In particular, the emergence of two distinct nonlinear branches in the pair-correlation embeddings provides a direct representation of the separation between weak-pairing and strong-pairing regimes. This distinction is not imposed through any predefined criterion but emerges automatically from the structure of the data illustrating the ability of nonlinear machine-learning techniques to reveal collective organization hidden in high-dimensional correlation datasets.
\subsection{Finite-Size Scaling}
\label{fss}
The results presented thus far consistently indicate an interaction-driven crossover from weakly correlated regime to strongly paired state. However, all numerical calculations are necessarily performed on finite cylinders, where finite-size effects can influence both energetic and correlation-based observables. It is therefore important to assess the extent to which the observed trends persist as the system size increases.

To this end, we perform finite-size scaling analysis using DMRG results obtained for $L_y\times4$ cylinders with $L_y=2$--$6$. The ground-state energy provides a particularly suitable quantity for such an analysis because it is a bulk thermodynamic observable that is less sensitive to local boundary effects than correlation functions. By examining the scaling of the ground-state energy per site with inverse cylinder width, we obtain estimates of the thermodynamic-limit behavior and evaluate the robustness of the interaction-driven crossover identified in the preceding sections. Since the leading finite-size correction to the ground-state energy is expected to scale inversely with the linear system size, the energy per site is fitted using the linear scaling form
\begin{equation}
\frac{E_0(L_y)}{M}=a+\frac{b}{L_y},
\label{eq:FSS_E0}
\end{equation}
where $M=4L_y$ is the total number of lattice sites, $a$ represents the extrapolated ground-state energy per site in the thermodynamic limit and $b$ characterizes the leading finite-size correction.

\begin{figure}[h]
\centering
\includegraphics[scale=0.37]{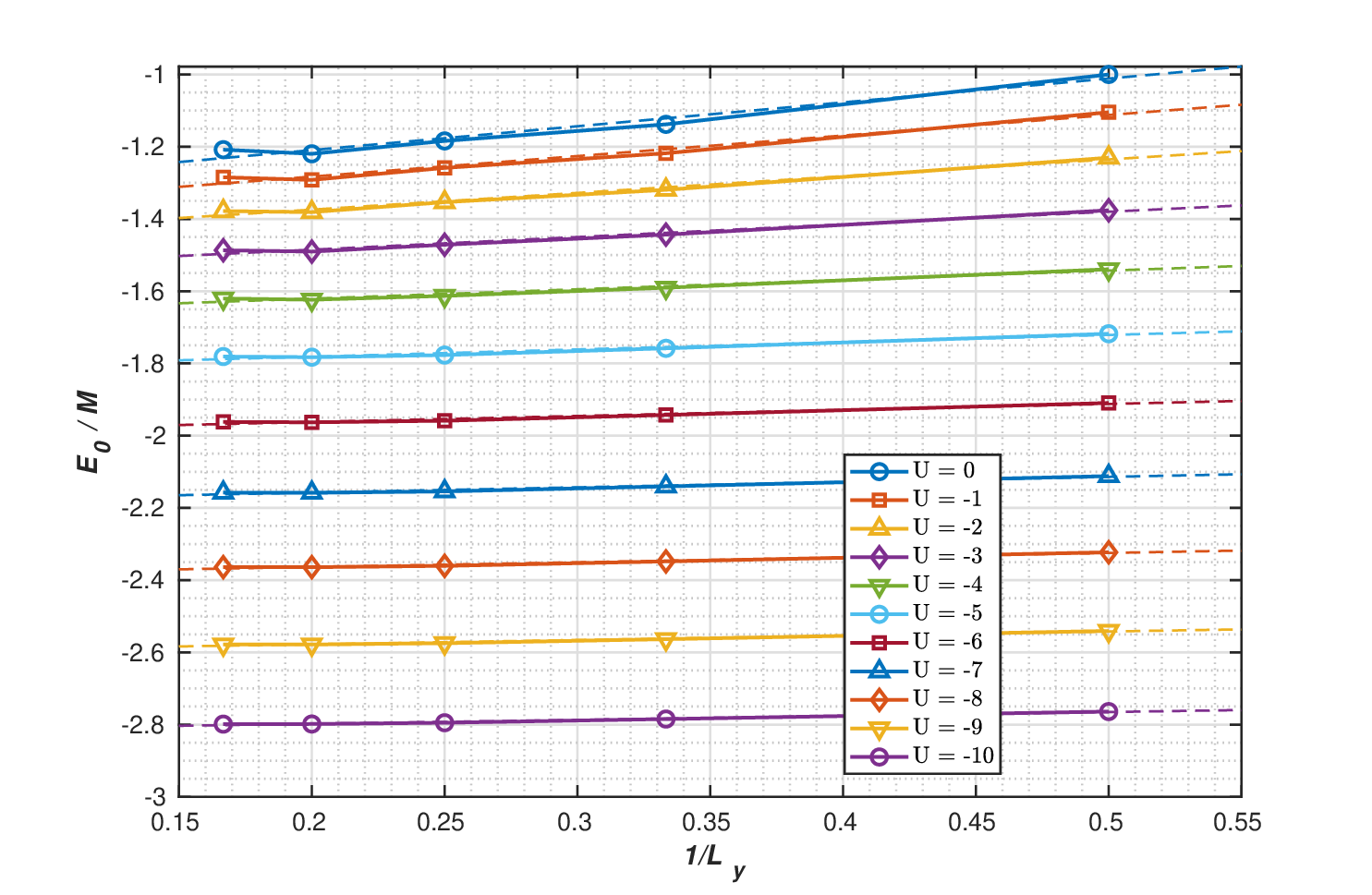}
\caption{Finite-size scaling of the ground-state energy per site $E_0/M$ as a function of $1/L_y$ for quarter-filled $L_y\times4$ cylinders at attractive interactions $U=0$ to $-10$. Symbols denote DMRG results, while dashed lines are linear fits to Eq.~(\ref{eq:FSS_E0}) used to extrapolate the ground-state energy to the thermodynamic limit ($1/L_y\rightarrow0$).}
\label{fig:Fig8}
\end{figure}
Figure \ref{fig:Fig8} shows the finite-size scaling of the ground-state energy per site for interaction strengths ranging from $U=0$ to $U=-10$. The DMRG results exhibit an approximately linear dependence on $1/L_y$ over the entire interaction range, demonstrating that the scaling form in Eq. (\ref{eq:FSS_E0}) provides an excellent description of the finite-size behavior. The dashed lines represent linear extrapolations to the thermodynamic limit ($1/L_y\rightarrow0$).
For all interaction strengths, the extrapolated ground-state energy decreases monotonically as the attractive interaction becomes stronger, reflecting the increasing stabilization of the many-body ground state through attractive pairing. This trend is consistent with the finite-cluster results presented in Sec. \ref{gspce}, where stronger attraction was found to enhance double occupancy, suppress local magnetic moments, strengthen pairing correlations and produce increasingly negative two-hole binding energies.

An additional feature evident from Fig. \ref{fig:Fig8} is that the finite-size dependence becomes progressively weaker with increasing attractive interaction. While the weak-coupling regime exhibits relatively larger size corrections, the strongly attractive regime displays nearly flat scaling curves, indicating that the ground-state energy rapidly approaches its thermodynamic-limit value. The absence of anomalous finite-size behavior or abrupt changes in the extrapolated energies further supports the picture of a continuous interaction-driven crossover rather than a sharp phase transition.

Overall, the finite-size scaling confirms that the interaction dependence of the ground-state energy obtained from finite cylinders remains robust upon extrapolation toward the thermodynamic limit, providing a reliable foundation for the subsequent finite-size analysis of other physical observables.

\begin{figure}[h]
\centering
\includegraphics[scale=0.37]{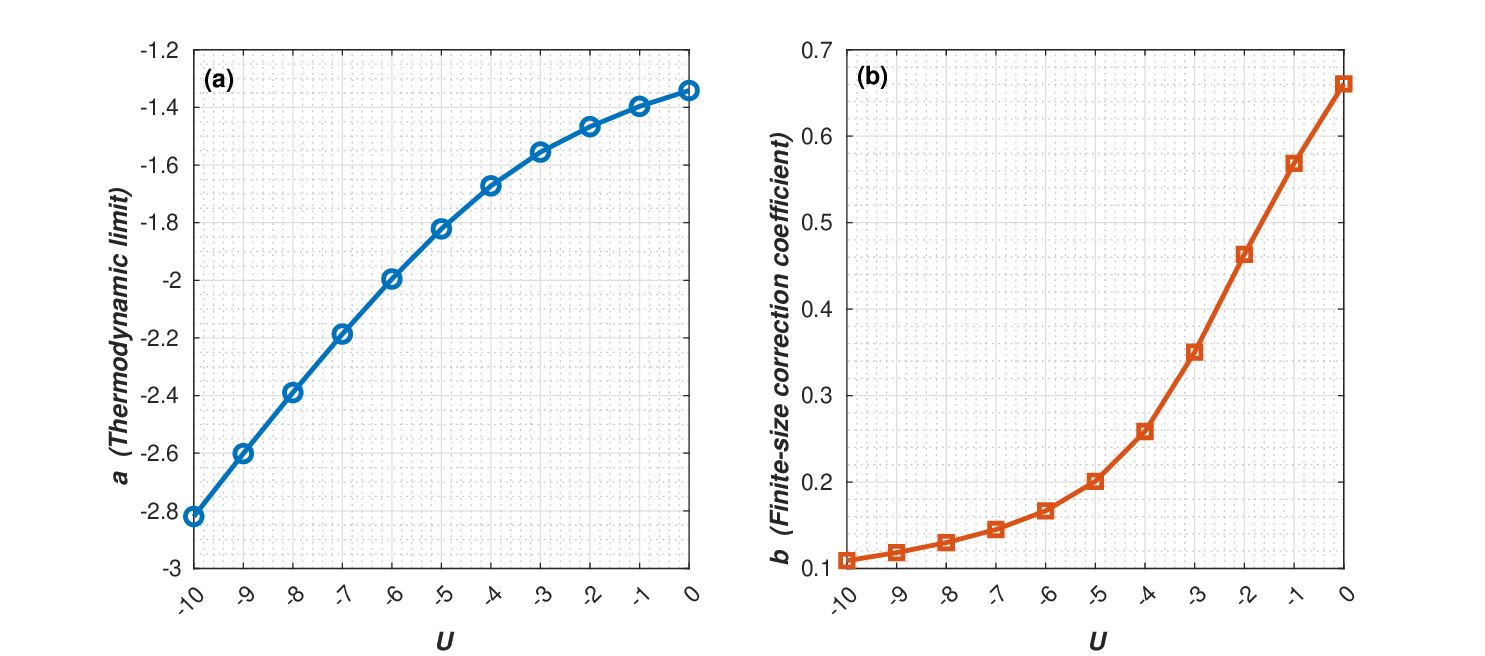}
\caption{Interaction dependence of the fitting parameters obtained from the finite-size scaling of the ground-state energy. (a) Thermodynamic-limit ground-state energy per site $a$ and (b) finite-size correction coefficient $b$, extracted from the linear fit $E_0(L_y)/M=a+b/L_y$. The monotonic evolution of both parameters indicates the continuous stabilization of the ground state and the progressive reduction of finite-size effects with increasing attractive interaction.}
\label{fig:Fig9}
\end{figure}
Figure \ref{fig:Fig9} summarizes the fitting parameters obtained from the finite-size scaling analysis of the ground-state energy using Eq. (\ref{eq:FSS_E0}). Panel (a) shows the extrapolated thermodynamic-limit ground-state energy per site $a$ while panel (b) presents the corresponding finite-size correction coefficient $b$ as functions of the attractive interaction strength $U$.
As shown in Fig. \ref{fig:Fig9}(a), the extrapolated ground-state energy decreases monotonically with increasing attractive interaction reflecting the progressive stabilization of the many-body ground state through pair formation. The decrease is most pronounced in the weak- and intermediate-coupling regimes whereas the variation becomes more gradual at stronger attraction indicating that the energy approaches its strong-coupling behavior smoothly. This trend is consistent with the finite-cluster results discussed previously and supports the continuous interaction-driven crossover identified from the local observables, correlation functions, binding energies and machine-learning analyses.

The finite-size correction coefficient shown in Fig. \ref{fig:Fig9}(b), remains positive over the entire interaction range but decreases systematically as the attractive interaction becomes stronger. This behavior indicates that finite-size effects are progressively reduced in the strong-coupling regime, where tightly bound local pairs lead to a more localized many-body state with weaker sensitivity to the system size. Consequently, the thermodynamic-limit energy is reached more rapidly for large negative $U$, consistent with the nearly flat finite-size scaling curves observed in Fig. \ref{fig:Fig8}. The smooth interaction dependence of both fitting parameters further reinforces the absence of any abrupt change in the ground-state properties across the attractive regime considered.

To further assess the robustness of the pairing correlations against finite-size effects, we perform finite-size scaling analysis of the singlet pairing structure factor $S_P(0,0)$. Since only a limited number of cylinder widths ($L_y=2$--$6$) are accessible within our DMRG calculations, we adopt the simplest scaling ansatz containing the leading finite-size correction,
\begin{equation}
S_P(0,0;L_y)=a+\frac{b}{L_y},
\label{eq:FSS_pair}
\end{equation}
where $a$ denotes the extrapolated thermodynamic-limit value and $b$ characterizes the leading finite-size correction. The resulting extrapolations should therefore be regarded as estimates of the thermodynamic trend rather than precise thermodynamic-limit determinations.
\begin{figure}[h]
\centering
\includegraphics[scale=0.37]{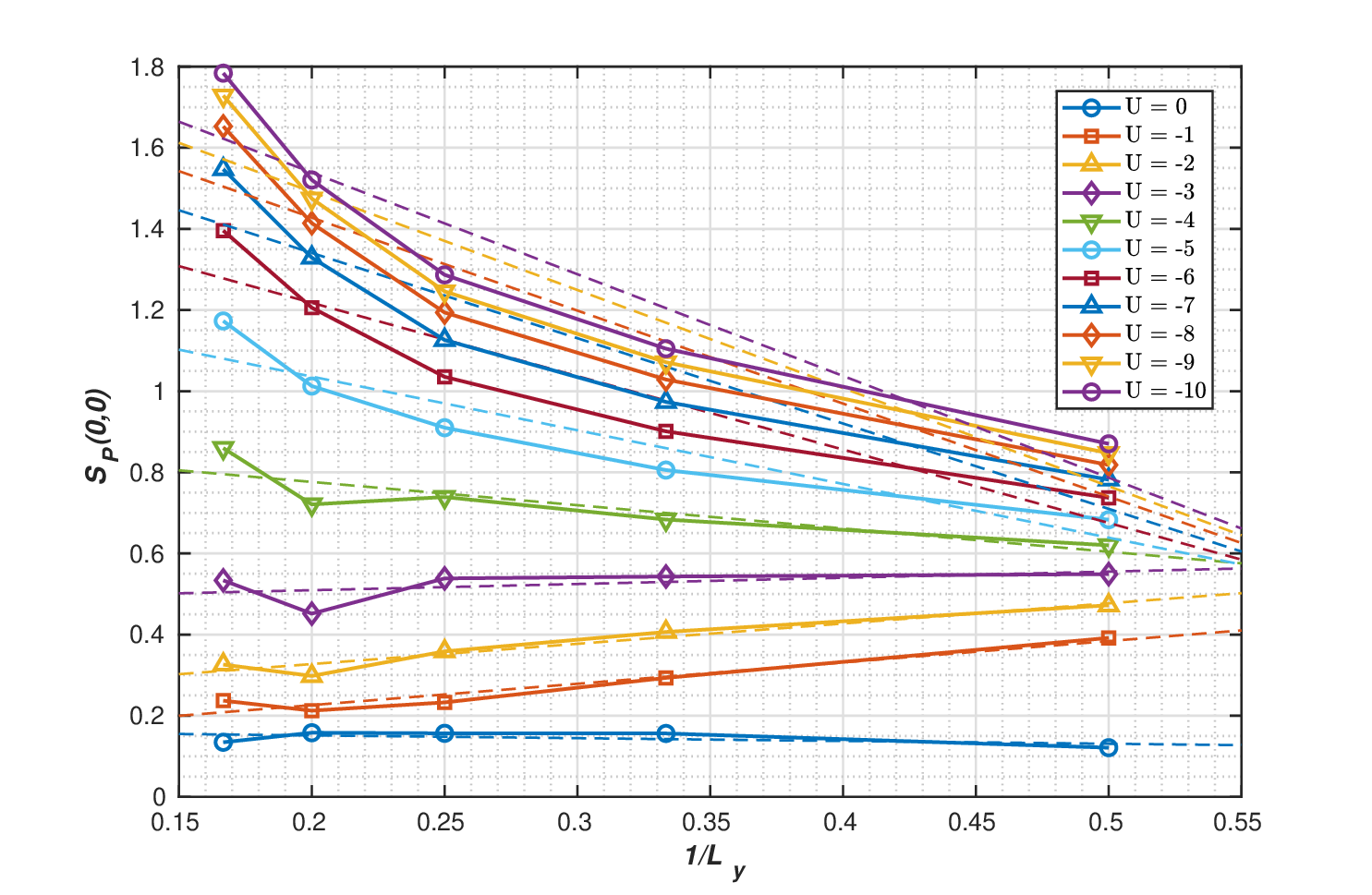}
\caption{Finite-size scaling of the singlet pairing structure factor $S_P(0,0)$ as a function of $1/L_y$ for quarter-filled $L_y\times4$ cylinders at attractive interactions $U=0$ to $-10$. Symbols denote DMRG results and dashed lines represent fits to the leading-order scaling form $S_P(0,0)=a+b/L_y$. The extrapolated values exhibit a systematic enhancement with increasing attraction indicating that the strengthening of pairing correlations persists beyond the finite cylinders considered here.}
\label{fig:Fig10}
\end{figure}

Figure \ref{fig:Fig10} shows the finite-size dependence of $S_P(0,0)$ for interaction strengths ranging from $U=0$ to $U=-10$. For all interaction strengths, the data exhibit a systematic size dependence that can be approximately captured by the leading-order scaling form defined in Eq. \ref{eq:FSS_pair}. Although some deviations from strict linearity are visible, particularly in the intermediate-coupling region where the crossover occurs, the extrapolations provide a useful estimate of the overall thermodynamic trend. The corresponding fitting parameters are summarized in Table \ref{tab:pairing_fss} in Appendix \ref{FSSPSF}. It is worth noting that the extrapolated pairing structure factor remains finite even at $U=0$. This finite value does not indicate superconducting long-range order but originates from short-range pair correlations already present in the noninteracting Fermi system. The interaction-driven enhancement of $S_P(0,0)$ with increasing attraction is therefore the physically relevant signature of pair formation.

The extrapolated thermodynamic-limit values increase monotonically with increasing attractive interaction demonstrating that the enhancement of pairing correlations persists beyond the finite cylinders studied here. The most rapid increase occurs in the intermediate-coupling region $U\approx -4$ to $-6$, consistent with the crossover region identified from the excitation gap, local observables, binding energies and machine-learning-based analyses.
Although the magnitude of the finite-size correction coefficient $b$ grows at strong attraction (Appendix \ref{FSSPSF}), the extrapolated values remain large and continue to increase smoothly with increasing negative $U$. The smooth interaction dependence of the extrapolated pairing structure factor indicates a continuous reorganization of the pairing correlations rather than an abrupt phase transition. Taken together, these results suggest that the strong enhancement of pairing correlations observed on finite cylinders is not merely a finite-size effect and remains robust as the system size increases.

The PCA results presented in Sec. \ref{pca} revealed that the leading explained-variance ratio $\tilde{\lambda}_1$ of the pair-pair correlation matrices increases systematically with increasing attractive interaction indicating that the correlation landscape becomes progressively dominated by a single collective fluctuation pattern. To examine whether this machine-learning-based signature remains robust with increasing system size, we perform finite-size scaling analysis of $\tilde{\lambda}_1$ obtained from the PCA of pair-pair correlation matrices for $L_y\times4$ cylinders with $L_y=2$--$6$.

As for the pairing structure factor, the limited range of accessible cylinder widths motivates the use of a leading-order finite-size scaling form
\begin{equation}
\tilde{\lambda}_1(L_y)=a+\frac{b}{L_y},
\label{eq:FSS_PCA}
\end{equation}
where $a$ represents the extrapolated thermodynamic-limit value and $b$ characterizes the leading finite-size correction. The extrapolated values should therefore be interpreted as indicators of the overall thermodynamic tendency rather than precise thermodynamic-limit estimates.

\begin{figure}[h]
\centering
\includegraphics[scale=0.37]{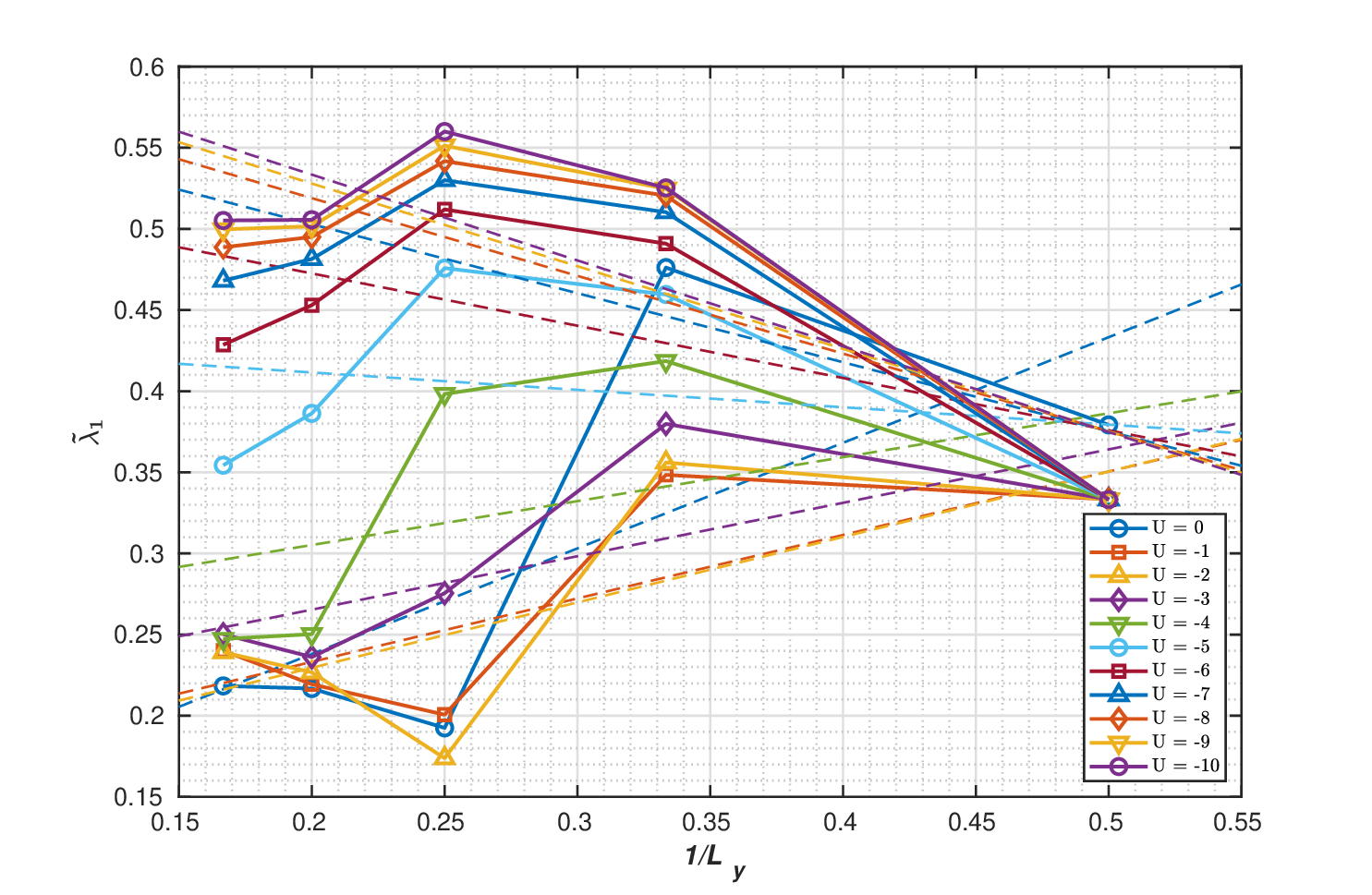}
\caption{Finite-size scaling of the leading explained-variance ratio $\tilde{\lambda}_1$ obtained from the PCA of pair-pair correlation matrices for quarter-filled $L_y\times4$ cylinders. Symbols denote DMRG results and dashed lines represent fits to the leading-order scaling form $\tilde{\lambda}_1(L_y)=a+b/L_y$. The extrapolated values display a pronounced increase across the intermediate-coupling region indicating that the pair-correlation landscape becomes increasingly dominated by a single collective fluctuation channel.}
\label{fig:Fig11}
\end{figure}
Figure \ref{fig:Fig11} shows the finite-size scaling of the leading explained-variance ratio $\tilde{\lambda}_1$. For each interaction strength, the finite-size dependence is analyzed using the leading-order scaling form of Eq. (\ref{eq:FSS_PCA}) and the corresponding fitting parameters are summarized in Table \ref{tab:pca_fss} in Appendix \ref{app:PCAFSS}. While deviations from strict linear behavior are visible for some interaction strengths particularly near the crossover region, the extrapolated values reveal clear and systematic interaction-dependent trends.

The extrapolated thermodynamic-limit value $a$ exhibits a pronounced interaction dependence. For weak attraction ($U=0$ to $-4$), $a$ remains small indicating that the variance of the pair-correlation data is distributed among several principal components. A rapid increase occurs between $U=-4$ and $U=-5$, after which $a$ continues to grow and reaches approximately $0.64$ at $U=-10$. This behavior indicates that the pair-correlation landscape becomes increasingly dominated by a single collective fluctuation pattern as the attractive interaction strengthens.

A corresponding change is observed in the finite-size correction coefficient $b$. While $b$ is positive in the weak-coupling regime, it becomes negative for $U\le -5$ (Appendix \ref{app:PCAFSS}), signaling a qualitative change in the finite-size evolution of the leading principal component. The crossover region identified from the finite-size-scaled PCA data therefore coincides with the interaction range previously inferred from the excitation gap, local observables, binding energies and UMAP analysis.

Physically, this behavior reflects the evolution from a weak-coupling regime characterized by extended Cooper-like pairing fluctuations to a strong-coupling regime dominated by tightly bound local singlet pairs. The increasing dominance of a single principal fluctuation channel beyond the crossover region indicates that the pair-correlation landscape becomes progressively more coherent and low dimensional. The persistence of this trend after extrapolation suggests that the machine-learning signatures of the crossover are not finite-size artifacts and remain consistent with the expected BCS--BEC-like evolution in the AHM.
\section{Summary and Conclusion}
\label{summary}
In this work, we have investigated the quarter-filled AHM on finite-width cylindrical lattices by combining ED, DMRG and unsupervised machine-learning techniques. This integrated approach enables a comprehensive characterization of the evolution of the many-body ground state from both conventional physical observables and unsupervised correlation analysis.

The DMRG ground-state energies are in excellent agreement with ED results, establishing the reliability of the numerical calculations. As the attractive interaction increases, the ground-state energy decreases monotonically while the excitation gap exhibits a broad maximum at intermediate coupling signaling a smooth crossover from weakly bound Cooper-like pairs to a regime dominated by tightly bound local singlet pairs. This crossover originates from the competition between kinetic-energy-driven itinerancy and interaction-driven onsite pair formation and represents the finite-size signatures of the BCS--BEC crossover in the thermodynamic limit, which is consistently supported by the monotonic increase of the double occupancy, suppression of the local magnetic moment, enhancement of the charge and pairing structure factors and the growth of real-space pair-pair correlations.
The hole-binding-energy analysis provides direct energetic evidence for pairing. The two-hole binding energy remains negative throughout the attractive regime demonstrating robust pair formation, whereas three-hole binding appears only at sufficiently strong attraction and exhibits pronounced finite-size dependence. In contrast, the four-hole binding energy remains positive over the entire parameter range indicating that the dominant instability is pair formation rather than extensive multi-hole clustering or phase separation.

Beyond these conventional observables, we employed PCA and UMAP to analyze the full correlation matrices without assuming any predefined order parameter. Unlike conventional quantities, which probe selected aspects of the many-body state, the machine-learning-based methods utilize all correlation information simultaneously and therefore provide a global characterization of the interaction-driven evolution. PCA reveals a systematic redistribution of the explained variance with increasing interaction strength: charge correlations evolve toward a more distributed multi-component representation whereas pair-pair correlations become increasingly dominated by a single principal fluctuation channel. UMAP further exposes the intrinsic nonlinear organization of the correlation data, with the pair-pair correlation matrices separating into distinct weak- and strong-coupling branches while the charge-correlation matrices evolve smoothly along a continuous manifold. Importantly, these signatures emerge directly from the data without imposing an order parameter or crossover criterion. The coincidence of the machine-learning signatures with the crossover region identified from the excitation gap, local observables and binding energies demonstrates that PCA and UMAP provide an objective and complementary route for detecting interaction-driven reorganization of the many-body correlation landscape.

Finally, finite-size scaling analyses of the ground-state energy, the pairing structure factor and the leading PCA variance ratio demonstrate that the interaction-driven trends remain robust with increasing cylinder width. The extrapolated thermodynamic-limit ground-state energy decreases smoothly with increasing attraction while the pairing structure factor exhibits a pronounced enhancement that persists beyond the finite systems considered. Moreover, the thermodynamic-limit value of the leading PCA variance ratio increases substantially across the intermediate-coupling region indicating that the pairing correlation matrices become progressively dominated by a single collective fluctuation channel. The consistent behavior of these independent quantities provides strong evidence that the crossover identified near $U\approx-4$ to $-6$ is not a finite-size artifact but a robust manifestation of the evolution from weakly overlapping Cooper pairs to tightly bound local pairs analogous to the BCS--BEC crossover expected in the thermodynamic limit.

Overall, the present work demonstrates that combining state-of-the-art many-body numerical methods with unsupervised machine-learning techniques provides a powerful framework for identifying interaction-driven crossovers in correlated quantum systems. The methodology developed here is general and can be readily extended to other strongly correlated lattice models, offering a promising route toward the unsupervised learning characterization of emergent quantum phases where conventional order parameters are absent or difficult to identify.
\section*{Appendix}
\appendix
\section{Finite-Size Scaling Parameters of the Pairing Structure Factor}
\label{FSSPSF}
The fitting parameters obtained from the linear extrapolation of the pairing structure factor are listed in Table \ref{tab:pairing_fss}. The thermodynamic-limit value $a$ increases monotonically with increasing attractive interaction, while the magnitude of the finite-size correction coefficient $b$ becomes larger in the strong-coupling regime. The persistence of large extrapolated values demonstrates that the enhancement of pairing correlations remains robust in the thermodynamic limit despite the increasing finite-size corrections.

\begin{table}[h]
\caption{Fitting parameters obtained from the finite-size scaling of the singlet pairing structure factor using $S_P(0,0)=a+b/L_y$. Here $a$ denotes the extrapolated thermodynamic-limit value and $b$ is the leading finite-size correction coefficient.}
\label{tab:pairing_fss}
\begin{ruledtabular}
\begin{tabular}{ccc}
$U$ & $a$ & $b$ \\
\hline
 0  & 0.165260 & -0.068650 \\
 \hline
-1  & 0.120903 &  0.526231 \\
\hline
-2  & 0.227486 &  0.499298 \\
\hline
-3  & 0.478820 &  0.152885 \\
\hline
-4  & 0.890885 & -0.573278 \\
\hline
-5  & 1.300667 & -1.323532 \\
\hline
-6  & 1.579547 & -1.808728 \\
\hline
-7  & 1.761318 & -2.102767 \\
\hline
-8  & 1.886115 & -2.291794 \\
\hline
-9  & 1.974797 & -2.418176 \\
\hline
-10 & 2.040243 & -2.506821 \\
\end{tabular}
\end{ruledtabular}
\end{table}
\section{Finite-Size Scaling Parameters of the Leading PCA Variance Ratio}
\label{app:PCAFSS}
The fitting parameters obtained from the finite-size scaling analysis of the leading explained-variance ratio $\tilde{\lambda}_1$ are summarized in Table \ref{tab:pca_fss}. The thermodynamic-limit value $a$ increases systematically with increasing attractive interaction indicating that the pair-correlation data become progressively dominated by a single principal fluctuation channel. A significant increase in $a$ occurs between $U=-4$ and $U=-5$, consistent with the crossover region identified throughout the main text. The finite-size correction coefficient $b$ simultaneously changes sign across the same interaction range providing additional evidence for a qualitative reorganization of the correlation landscape. These results demonstrate that the PCA signatures of the crossover remain robust after extrapolation toward the thermodynamic limit.
\begin{table}[h]
\caption{Fitting parameters obtained from the finite-size scaling of the leading explained-variance ratio using $\tilde{\lambda}_1(L_y)=a+b/L_y$. Here $a$ denotes the extrapolated thermodynamic-limit value of the leading variance ratio and $b$ is the leading finite-size correction coefficient.}
\label{tab:pca_fss}
\begin{ruledtabular}
\begin{tabular}{ccc}
$U$ & $a$ & $b$ \\
\hline
 0  & 0.107885 &  0.650740 \\
\hline
-1  & 0.154963 &  0.391051 \\
\hline
-2  & 0.148880 &  0.403196 \\
\hline
-3  & 0.199556 &  0.329109 \\
\hline
-4  & 0.251095 &  0.270535 \\
\hline
-5  & 0.432943 & -0.107131 \\
\hline
-6  & 0.536991 & -0.322197 \\
\hline
-7  & 0.587988 & -0.425411 \\
\hline
-8  & 0.614756 & -0.479010 \\
\hline
-9  & 0.629836 & -0.509412 \\
\hline
-10 & 0.639184 & -0.528718 \\
\end{tabular}
\end{ruledtabular}
\end{table}
\bibliography{manuscript}

@article{Hubbard1963,
    author = {Hubbard, J.},
    title = {Electron correlations in narrow energy bands},
    journal = {Proceedings of the Royal Society of London. A. Mathematical and Physical Sciences},
    volume = {276},
    number = {1365},
    pages = {238-257},
    year = {1963},
    month = {11},
    issn = {0080-4630},
    doi = {10.1098/rspa.1963.0204},
    url = {https://doi.org/10.1098/rspa.1963.0204}
}

@article{Dagotto1994,
  title = {Correlated electrons in high-temperature superconductors},
  author = {Dagotto, Elbio},
  journal = {Rev. Mod. Phys.},
  volume = {66},
  issue = {3},
  pages = {763--840},
  numpages = {0},
  year = {1994},
  month = {Jul},
  publisher = {American Physical Society},
  doi = {10.1103/RevModPhys.66.763},
  url = {https://link.aps.org/doi/10.1103/RevModPhys.66.763}
}

@article{Daniel2022,
   author = "Arovas, Daniel P. and Berg, Erez and Kivelson, Steven A. and Raghu, Srinivas",
   title = "The Hubbard Model", 
   journal= "Annual Review of Condensed Matter Physics",
   year = "2022",
   volume = "13",
   number = "Volume 13, 2022",
   pages = "239-274",
   doi = "https://doi.org/10.1146/annurev-conmatphys-031620-102024",
   url = "https://www.annualreviews.org/content/journals/10.1146/annurev-conmatphys-031620-102024",
   publisher = "Annual Reviews",
   issn = "1947-5462",
   type = "Journal Article",
}

@article{Mott1974,
author = {L. M. Falicov},
title = {The Mott Transition: Metal-Insulator Transitions, N. F. Mott. Taylor and Francis, London, and Barnes and Noble, New York, 1974, xvi, 278 pp.},
journal = {Science},
volume = {188},
number = {4192},
pages = {1007-1008},
year = {1975},
doi = {10.1126/science.188.4192.1007.b},
URL = {https://www.science.org/doi/abs/10.1126/science.188.4192.1007.b}
}

@article{Gutzwiller1963,
  title = {Effect of Correlation on the Ferromagnetism of Transition Metals},
  author = {Gutzwiller, Martin C.},
  journal = {Phys. Rev. Lett.},
  volume = {10},
  issue = {5},
  pages = {159--162},
  numpages = {0},
  year = {1963},
  month = {Mar},
  publisher = {American Physical Society},
  doi = {10.1103/PhysRevLett.10.159},
  url = {https://link.aps.org/doi/10.1103/PhysRevLett.10.159}
}

@ARTICLE{Kanamori1963,
       author = {{Kanamori}, J.},
        title = "{Electron Correlation and Ferromagnetism of Transition Metals}",
      journal = {Progress of Theoretical Physics},
         year = 1963,
        month = sep,
       volume = {30},
       number = {3},
        pages = {275-289},
          doi = {10.1143/PTP.30.275},
       adsurl = {https://ui.adsabs.harvard.edu/abs/1963PThPh..30..275K}
}

@article{Kaito2022,
  title = {Skyrmion and vortex crystals in the Hubbard model},
  author = {Kobayashi, Kaito and Hayami, Satoru},
  journal = {Phys. Rev. B},
  volume = {106},
  issue = {14},
  pages = {L140406},
  numpages = {6},
  year = {2022},
  month = {Oct},
  publisher = {American Physical Society},
  doi = {10.1103/PhysRevB.106.L140406},
  url = {https://link.aps.org/doi/10.1103/PhysRevB.106.L140406}
}

@article{Fahadboh2025,
  title={Binding of holes and competing spin-charge order in simple and extended Hubbard model on cylindrical lattice: An exact diagonalization study},
  author={Equbal, Md Fahad and Ahsan, M. A. H},
  journal={arXiv preprint arXiv:2512.10577},
  url = {https://doi.org/10.48550/arXiv.2512.10577},
  year={2025}
}

@article{Fahadalm2026,
  title={Emergence of correlation-driven altermagnetism in Hubbard model on geometrically frustrated lattice-clusters},
  author={Equbal, Md Fahad and Ahsan, M. A. H},
  journal={arXiv preprint arXiv:2603.00536},
  url = {https://doi.org/10.48550/arXiv.2603.00536},
  year={2026}
}

@article{Micnas1990,
  title = {Superconductivity in narrow-band systems with local nonretarded attractive interactions},
  author = {Micnas, R. and Ranninger, J. and Robaszkiewicz, S.},
  journal = {Rev. Mod. Phys.},
  volume = {62},
  issue = {1},
  pages = {113--171},
  numpages = {0},
  year = {1990},
  month = {Jan},
  publisher = {American Physical Society},
  doi = {10.1103/RevModPhys.62.113},
  url = {https://link.aps.org/doi/10.1103/RevModPhys.62.113}
}

@article{Chen2024,
  title = {When superconductivity crosses over: From BCS to BEC},
  author = {Chen, Qijin and Wang, Zhiqiang and Boyack, Rufus and Yang, Shuolong and Levin, K.},
  journal = {Rev. Mod. Phys.},
  volume = {96},
  issue = {2},
  pages = {025002},
  numpages = {54},
  year = {2024},
  month = {May},
  publisher = {American Physical Society},
  doi = {10.1103/RevModPhys.96.025002},
  url = {https://link.aps.org/doi/10.1103/RevModPhys.96.025002}
}

@article{Chan2020,
  title = {Pair correlations in the attractive Hubbard model},
  author = {Chan, C. F. and Gall, M. and Wurz, N. and K\"ohl, M.},
  journal = {Phys. Rev. Res.},
  volume = {2},
  issue = {2},
  pages = {023210},
  numpages = {6},
  year = {2020},
  month = {May},
  publisher = {American Physical Society},
  doi = {10.1103/PhysRevResearch.2.023210},
  url = {https://link.aps.org/doi/10.1103/PhysRevResearch.2.023210}
}

@article{Zhu2025,
  title = {Rigorous demonstration of pair-density-wave superconductivity in the ${\ensuremath{\sigma}}_{z}$-Hubbard model},
  author = {Zhu, Xingchuan and Sun, Junsong and Gong, Shou-Shu and Huang, Wen and Feng, Shiping and Scalettar, Richard T. and Guo, Huaiming},
  journal = {Phys. Rev. B},
  volume = {111},
  issue = {4},
  pages = {045158},
  numpages = {6},
  year = {2025},
  month = {Jan},
  publisher = {American Physical Society},
  doi = {10.1103/PhysRevB.111.045158},
  url = {https://link.aps.org/doi/10.1103/PhysRevB.111.045158}
}

@article{He2025,
  title = {$\ensuremath{\eta}$-pairing states in the Hubbard model with nonuniform Hubbard interaction},
  author = {He, D. K. and Song, Z.},
  journal = {Phys. Rev. B},
  volume = {112},
  issue = {7},
  pages = {075135},
  numpages = {7},
  year = {2025},
  month = {Aug},
  publisher = {American Physical Society},
  doi = {10.1103/h3yg-lpf8},
  url = {https://link.aps.org/doi/10.1103/h3yg-lpf8}
}

@article{Ho2009,
  title = {Quantum simulation of the Hubbard model: The attractive route},
  author = {Ho, A. F. and Cazalilla, M. A. and Giamarchi, T.},
  journal = {Phys. Rev. A},
  volume = {79},
  issue = {3},
  pages = {033620},
  numpages = {11},
  year = {2009},
  month = {Mar},
  publisher = {American Physical Society},
  doi = {10.1103/PhysRevA.79.033620},
  url = {https://link.aps.org/doi/10.1103/PhysRevA.79.033620}
}

@article{Titas2025,
doi = {10.1088/1361-6633/adc3a7},
url = {https://doi.org/10.1088/1361-6633/adc3a7},
year = {2025},
month = {apr},
publisher = {IOP Publishing},
volume = {88},
number = {4},
pages = {044501},
author = {Chanda, Titas and Barbiero, Luca and Lewenstein, Maciej and Mark, Manfred J and Zakrzewski, Jakub},
title = {Recent progress on quantum simulations of non-standard Bose–Hubbard models},
journal = {Reports on Progress in Physics}
}

@article{Kaneko2014,
author = {Kaneko ,Tatsuya and Ohta ,Yukinori},
title = {BCS–BEC Crossover in the Two-Dimensional Attractive Hubbard Model: Variational Cluster Approach},
journal = {Journal of the Physical Society of Japan},
volume = {83},
number = {2},
pages = {024711},
year = {2014},
doi = {10.7566/JPSJ.83.024711},
URL = { https://doi.org/10.7566/JPSJ.83.024711}
}

@article{Rodrigo2022,
  title = {Two-dimensional attractive Hubbard model and the BCS-BEC crossover},
  author = {Fontenele, Rodrigo A. and Costa, Natanael C. and dos Santos, Raimundo R. and Paiva, Thereza},
  journal = {Phys. Rev. B},
  volume = {105},
  issue = {18},
  pages = {184502},
  numpages = {9},
  year = {2022},
  month = {May},
  publisher = {American Physical Society},
  doi = {10.1103/PhysRevB.105.184502},
  url = {https://link.aps.org/doi/10.1103/PhysRevB.105.184502}
}

@article{Tong2024,
  title = {Disentangling the physics of the attractive Hubbard model as a fully interacting model of fermions via the accessible and symmetry-resolved entanglement entropies},
  author = {Shen, Tong and Barghathi, Hatem and Del Maestro, Adrian and Rubenstein, Brenda M.},
  journal = {Phys. Rev. B},
  volume = {109},
  issue = {19},
  pages = {195119},
  numpages = {18},
  year = {2024},
  month = {May},
  publisher = {American Physical Society},
  doi = {10.1103/PhysRevB.109.195119},
  url = {https://link.aps.org/doi/10.1103/PhysRevB.109.195119}
}

@article{Scalettar1989,
  title = {Phase diagram of the two-dimensional negative-U Hubbard model},
  author = {Scalettar, R. T. and Loh, E. Y. and Gubernatis, J. E. and Moreo, A. and White, S. R. and Scalapino, D. J. and Sugar, R. L. and Dagotto, E.},
  journal = {Phys. Rev. Lett.},
  volume = {62},
  issue = {12},
  pages = {1407--1410},
  numpages = {0},
  year = {1989},
  month = {Mar},
  publisher = {American Physical Society},
  doi = {10.1103/PhysRevLett.62.1407},
  url = {https://link.aps.org/doi/10.1103/PhysRevLett.62.1407}
}

@article{Seher2018,
  title = {Study of the superconducting order parameter in the two-dimensional negative-$U$ Hubbard model by grand-canonical twist-averaged boundary conditions},
  author = {Karakuzu, Seher and Seki, Kazuhiro and Sorella, Sandro},
  journal = {Phys. Rev. B},
  volume = {98},
  issue = {7},
  pages = {075156},
  numpages = {10},
  year = {2018},
  month = {Aug},
  publisher = {American Physical Society},
  doi = {10.1103/PhysRevB.98.075156},
  url = {https://link.aps.org/doi/10.1103/PhysRevB.98.075156}
}

@article{Fahad2025,
doi = {10.1088/1402-4896/adb45f},
url = {https://doi.org/10.1088/1402-4896/adb45f},
year = {2025},
month = {feb},
publisher = {IOP Publishing},
volume = {100},
number = {3},
pages = {035964},
author = {Equbal, Md Fahad and Hassan, S R and Ahsan, M A H},
title = {Exact diagonalization study of ground state properties and level statistics in simple and extended Hubbard lattice cluster},
journal = {Physica Scripta}
}

@article{Leticia2018,
     author = {Leticia Tarruell and Laurent Sanchez-Palencia},
     title = {Quantum simulation of the {Hubbard} model with ultracold fermions in optical lattices},
     journal = {Comptes Rendus. Physique},
     pages = {365--393},
     year = {2018},
     publisher = {Elsevier},
     volume = {19},
     number = {6},
     doi = {10.1016/j.crhy.2018.10.013},
     url = {https://doi.org/10.1016/j.crhy.2018.10.013}
}

@article{Cheng2010,
  title = {Feshbach resonances in ultracold gases},
  author = {Chin, Cheng and Grimm, Rudolf and Julienne, Paul and Tiesinga, Eite},
  journal = {Rev. Mod. Phys.},
  volume = {82},
  issue = {2},
  pages = {1225--1286},
  numpages = {0},
  year = {2010},
  month = {Apr},
  publisher = {American Physical Society},
  doi = {10.1103/RevModPhys.82.1225},
  url = {https://link.aps.org/doi/10.1103/RevModPhys.82.1225}
}

@article{Boll2016,
author = {Martin Boll  and Timon A. Hilker  and Guillaume Salomon  and Ahmed Omran  and Jacopo Nespolo  and Lode Pollet  and Immanuel Bloch  and Christian Gross },
title = {Spin- and density-resolved microscopy of antiferromagnetic correlations in Fermi-Hubbard chains},
journal = {Science},
volume = {353},
number = {6305},
pages = {1257-1260},
year = {2016},
doi = {10.1126/science.aag1635},
URL = {https://www.science.org/doi/abs/10.1126/science.aag1635}
}

@article{Mark2025,
  title = {Efficiently Measuring $d$-Wave Pairing and Beyond in Quantum Gas Microscopes},
  author = {Mark, Daniel K. and Hu, Hong-Ye and Kwan, Joyce and Kokail, Christian and Choi, Soonwon and Yelin, Susanne F.},
  journal = {Phys. Rev. Lett.},
  volume = {135},
  issue = {12},
  pages = {123402},
  numpages = {10},
  year = {2025},
  month = {Sep},
  publisher = {American Physical Society},
  doi = {10.1103/dqyf-kl8x},
  url = {https://link.aps.org/doi/10.1103/dqyf-kl8x}
}

@article{Gross2017,
author = {Christian Gross  and Immanuel Bloch },
title = {Quantum simulations with ultracold atoms in optical lattices},
journal = {Science},
volume = {357},
number = {6355},
pages = {995-1001},
year = {2017},
doi = {10.1126/science.aal3837},
URL = {https://www.science.org/doi/abs/10.1126/science.aal3837}
}

@article{Masaya2024,
  title = {Exact eigenstates of multicomponent Hubbard models: SU($N$) magnetic $\ensuremath{\eta}$ pairing, weak ergodicity breaking, and partial integrability},
  author = {Nakagawa, Masaya and Katsura, Hosho and Ueda, Masahito},
  journal = {Phys. Rev. Res.},
  volume = {6},
  issue = {4},
  pages = {043259},
  numpages = {31},
  year = {2024},
  month = {Dec},
  publisher = {American Physical Society},
  doi = {10.1103/PhysRevResearch.6.043259},
  url = {https://link.aps.org/doi/10.1103/PhysRevResearch.6.043259}
}

@article{Alexander2025,
author = {Alexander Guthmann  and Felix Lang  and Louisa Marie Kienesberger  and Sian Barbosa  and Artur Widera },
title = {Floquet engineering of Feshbach resonances in ultracold gases},
journal = {Science Advances},
volume = {11},
number = {40},
pages = {eadw3856},
year = {2025},
doi = {10.1126/sciadv.adw3856},
URL = {https://www.science.org/doi/abs/10.1126/sciadv.adw3856}
}

@article{Hirsch1985,
  title = {Two-dimensional Hubbard model: Numerical simulation study},
  author = {Hirsch, J. E.},
  journal = {Phys. Rev. B},
  volume = {31},
  issue = {7},
  pages = {4403--4419},
  numpages = {0},
  year = {1985},
  month = {Apr},
  publisher = {American Physical Society},
  doi = {10.1103/PhysRevB.31.4403},
  url = {https://link.aps.org/doi/10.1103/PhysRevB.31.4403}
}

@article{White1989,
  title = {Numerical study of the two-dimensional Hubbard model},
  author = {White, S. R. and Scalapino, D. J. and Sugar, R. L. and Loh, E. Y. and Gubernatis, J. E. and Scalettar, R. T.},
  journal = {Phys. Rev. B},
  volume = {40},
  issue = {1},
  pages = {506--516},
  numpages = {0},
  year = {1989},
  month = {Jul},
  publisher = {American Physical Society},
  doi = {10.1103/PhysRevB.40.506},
  url = {https://link.aps.org/doi/10.1103/PhysRevB.40.506}
}

@article{Chen2023,
  title = {Superconducting phases of the square-lattice extended Hubbard model},
  author = {Chen, Wei-Chih and Wang, Yao and Chen, Cheng-Chien},
  journal = {Phys. Rev. B},
  volume = {108},
  issue = {6},
  pages = {064514},
  numpages = {13},
  year = {2023},
  month = {Aug},
  publisher = {American Physical Society},
  doi = {10.1103/PhysRevB.108.064514},
  url = {https://link.aps.org/doi/10.1103/PhysRevB.108.064514}
}

@article{Callaway1990,
  title = {Small-cluster calculations for the simple and extended Hubbard models},
  author = {Callaway, J. and Chen, D. P. and Kanhere, D. G. and Li, Qiming},
  journal = {Phys. Rev. B},
  volume = {42},
  issue = {1},
  pages = {465--474},
  numpages = {0},
  year = {1990},
  month = {Jul},
  publisher = {American Physical Society},
  doi = {10.1103/PhysRevB.42.465},
  url = {https://link.aps.org/doi/10.1103/PhysRevB.42.465}
}

@article{Vijay2008,
  title = {Phase diagram of the attractive Hubbard model with inhomogeneous interactions},
  author = {Shenoy, Vijay B.},
  journal = {Phys. Rev. B},
  volume = {78},
  issue = {13},
  pages = {134503},
  numpages = {8},
  year = {2008},
  month = {Oct},
  publisher = {American Physical Society},
  doi = {10.1103/PhysRevB.78.134503},
  url = {https://link.aps.org/doi/10.1103/PhysRevB.78.134503}
}

@article{Toschi2005,
  title = {Energetic balance of the superconducting transition across the BCS---Bose Einstein crossover in the attractive Hubbard model},
  author = {Toschi, A. and Capone, M. and Castellani, C.},
  journal = {Phys. Rev. B},
  volume = {72},
  issue = {23},
  pages = {235118},
  numpages = {10},
  year = {2005},
  month = {Dec},
  publisher = {American Physical Society},
  doi = {10.1103/PhysRevB.72.235118},
  url = {https://link.aps.org/doi/10.1103/PhysRevB.72.235118}
}

@article{Jiang2019,
author = {Hong-Chen Jiang  and Thomas P. Devereaux },
title = {Superconductivity in the doped Hubbard model and its interplay with next-nearest hopping},
journal = {Science},
volume = {365},
number = {6460},
pages = {1424-1428},
year = {2019},
doi = {10.1126/science.aal5304},
URL = {https://www.science.org/doi/abs/10.1126/science.aal5304}
}

@article{Jiang2020,
  title = {Ground state phase diagram of the doped Hubbard model on the four-leg cylinder},
  author = {Jiang, Yi-Fan and Zaanen, Jan and Devereaux, Thomas P. and Jiang, Hong-Chen},
  journal = {Phys. Rev. Res.},
  volume = {2},
  issue = {3},
  pages = {033073},
  numpages = {14},
  year = {2020},
  month = {Jul},
  publisher = {American Physical Society},
  doi = {10.1103/PhysRevResearch.2.033073},
  url = {https://link.aps.org/doi/10.1103/PhysRevResearch.2.033073}
}

@article{Jiang2021,
author = {Hong-Chen Jiang  and Steven A. Kivelson },
title = {Stripe order enhanced superconductivity in the Hubbard model},
journal = {Proceedings of the National Academy of Sciences},
volume = {119},
number = {1},
pages = {e2109406119},
year = {2022},
doi = {10.1073/pnas.2109406119},
URL = {https://www.pnas.org/doi/abs/10.1073/pnas.2109406119}
}

@article{Jiang2024,
  title = {Ground-state phase diagram and superconductivity of the doped Hubbard model on six-leg square cylinders},
  author = {Jiang, Yi-Fan and Devereaux, Thomas P. and Jiang, Hong-Chen},
  journal = {Phys. Rev. B},
  volume = {109},
  issue = {8},
  pages = {085121},
  numpages = {6},
  year = {2024},
  month = {Feb},
  publisher = {American Physical Society},
  doi = {10.1103/PhysRevB.109.085121},
  url = {https://link.aps.org/doi/10.1103/PhysRevB.109.085121}
}

@article{Leiwang2016,
  title = {Discovering phase transitions with unsupervised learning},
  author = {Wang, Lei},
  journal = {Phys. Rev. B},
  volume = {94},
  issue = {19},
  pages = {195105},
  numpages = {5},
  year = {2016},
  month = {Nov},
  publisher = {American Physical Society},
  doi = {10.1103/PhysRevB.94.195105},
  url = {https://link.aps.org/doi/10.1103/PhysRevB.94.195105}
}

@article{Hu2017,
  title = {Discovering phases, phase transitions, and crossovers through unsupervised machine learning: A critical examination},
  author = {Hu, Wenjian and Singh, Rajiv R. P. and Scalettar, Richard T.},
  journal = {Phys. Rev. E},
  volume = {95},
  issue = {6},
  pages = {062122},
  numpages = {14},
  year = {2017},
  month = {Jun},
  publisher = {American Physical Society},
  doi = {10.1103/PhysRevE.95.062122},
  url = {https://link.aps.org/doi/10.1103/PhysRevE.95.062122}
}

@article{Costa2017,
  title = {Principal component analysis for fermionic critical points},
  author = {Costa, Natanael C. and Hu, Wenjian and Bai, Z. J. and Scalettar, Richard T. and Singh, Rajiv R. P.},
  journal = {Phys. Rev. B},
  volume = {96},
  issue = {19},
  pages = {195138},
  numpages = {10},
  year = {2017},
  month = {Nov},
  publisher = {American Physical Society},
  doi = {10.1103/PhysRevB.96.195138},
  url = {https://link.aps.org/doi/10.1103/PhysRevB.96.195138}
}

@article{Kiwata2019,
  title = {Deriving the order parameters of a spin-glass model using principal component analysis},
  author = {Kiwata, Hirohito},
  journal = {Phys. Rev. E},
  volume = {99},
  issue = {6},
  pages = {063304},
  numpages = {9},
  year = {2019},
  month = {Jun},
  publisher = {American Physical Society},
  doi = {10.1103/PhysRevE.99.063304},
  url = {https://link.aps.org/doi/10.1103/PhysRevE.99.063304}
}

@article{Leland2018,
  title={UMAP: Uniform Manifold Approximation and Projection for Dimension Reduction},
  author={McInnes, Leland and Healy, John and Melville, James},
  journal={arXiv preprint arXiv:1802.03426},
  url = {https://doi.org/10.48550/arXiv.1802.03426},
  year={2020}
}

@article{Joana2022,
doi = {10.1088/1742-6596/2407/1/012043},
url = {https://doi.org/10.1088/1742-6596/2407/1/012043},
year = {2022},
month = {dec},
publisher = {IOP Publishing},
volume = {2407},
number = {1},
pages = {012043},
author = {Teixeira, Joana and Rocha, Vicente and Oliveira, João and Jorge, Pedro A. S. and Silva, Nuno A.},
title = {Towards real-time identification of trapped particles with UMAP-based classifiers},
journal = {Journal of Physics: Conference Series}
}

@article{Khatami2019,
doi = {10.1088/1742-6596/1290/1/012006},
url = {https://doi.org/10.1088/1742-6596/1290/1/012006},
year = {2019},
month = {oct},
publisher = {IOP Publishing},
volume = {1290},
number = {1},
pages = {012006},
author = {Khatami, Ehsan},
title = {Principal component analysis of the magnetic transition in the three-dimensional Fermi-Hubbard model},
journal = {Journal of Physics: Conference Series}
}

@article{Fahadps2026,
doi = {10.1088/1402-4896/ae6ea0},
url = {https://doi.org/10.1088/1402-4896/ae6ea0},
year = {2026},
month = {may},
publisher = {IOP Publishing},
volume = {101},
number = {21},
pages = {215921},
author = {Fahad Equbal, Md and Hassan, S R and Ahsan, M A H},
title = {Principal component analysis of competing correlations in quarter-filled Hubbard models},
journal = {Physica Scripta}
}

@article{Fahadap2026,
title = {Spin-resolved Mott crossover and entanglement in the half-filled Hubbard model},
journal = {Annals of Physics},
volume = {488},
pages = {170414},
year = {2026},
issn = {0003-4916},
doi = {https://doi.org/10.1016/j.aop.2026.170414},
url = {https://www.sciencedirect.com/science/article/pii/S0003491626000734},
author = {Md Fahad Equbal and M. A. H. Ahsan}
}

@article{Sarmahsan1996,
author = {Sarma, C. R. and Ahsan, M. A. H.},
title = {Electron correlation studies: Rumer basis approach},
journal = {International Journal of Quantum Chemistry},
volume = {60},
number = {1},
pages = {147-156},
doi = {https://doi.org/10.1002/(SICI)1097-461X(1996)60:1<147::AID-QUA16>3.0.CO;2-C},
url = {https://onlinelibrary.wiley.com/doi/abs/10.1002/%28SICI%291097-461X%281996%2960%3A1%3C147%3A%3AAID-QUA16%3E3.0.CO%3B2-C},
year = {1996}
}

@article{Alvarez2009,
title = {The density matrix renormalization group for strongly correlated electron systems: A generic implementation},
journal = {Computer Physics Communications},
volume = {180},
number = {9},
pages = {1572-1578},
year = {2009},
issn = {0010-4655},
doi = {https://doi.org/10.1016/j.cpc.2009.02.016},
url = {https://www.sciencedirect.com/science/article/pii/S001046550900071X},
author = {G. Alvarez}
}

@article{Alvarez2011,
  title = {Time evolution with the density-matrix renormalization-group algorithm: A generic implementation for strongly correlated electronic systems},
  author = {Alvarez, G. and Dias da Silva, Luis G. G. V. and Ponce, E. and Dagotto, E.},
  journal = {Phys. Rev. E},
  volume = {84},
  issue = {5},
  pages = {056706},
  numpages = {7},
  year = {2011},
  month = {Nov},
  publisher = {American Physical Society},
  doi = {10.1103/PhysRevE.84.056706},
  url = {https://link.aps.org/doi/10.1103/PhysRevE.84.056706}
}

@article{Alvarez2012,
title = {Implementation of the SU(2) Hamiltonian symmetry for the DMRG algorithm},
journal = {Computer Physics Communications},
volume = {183},
number = {10},
pages = {2226-2232},
year = {2012},
issn = {0010-4655},
doi = {https://doi.org/10.1016/j.cpc.2012.04.025},
url = {https://www.sciencedirect.com/science/article/pii/S0010465512001646},
author = {Gonzalo Alvarez}
}

@article{Alvarez2013,
  title = {Production of minimally entangled typical thermal states with the Krylov-space approach},
  author = {Alvarez, G.},
  journal = {Phys. Rev. B},
  volume = {87},
  issue = {24},
  pages = {245130},
  numpages = {6},
  year = {2013},
  month = {Jun},
  publisher = {American Physical Society},
  doi = {10.1103/PhysRevB.87.245130},
  url = {https://link.aps.org/doi/10.1103/PhysRevB.87.245130}
}
\end{document}